\documentclass[intlimits,twoside,a4paper]{article}
\usepackage[cp1251]{inputenc}

\usepackage[eqsecnum]{cmpj3}
\usepackage{bm}% bold math
\issue{2025}{28}{1}{13602}
\doinumber{10.5488/CMP.28.13602}

\title[Water-methanol liquid mixtures]
{On the temperature, pressure and composition effects 
in the properties of water-methanol mixtures.
I. Density, excess mixing volume and enthalpy,  and self-diffusion
coefficients from molecular dynamics simulations
}

\author[M. Cruz Sanchez, V. Trejos Montoya, O. Pizio]
{M. Cruz Sanchez\orcid{0009-0005-0407-6496}\refaddr{label1},
V. Trejos Montoya\orcid{0000-0002-1136-3806}\refaddr{label1}, 
O. Pizio\orcid{0000-0001-8333-4652}\refaddr{label2}\thanks{Corresponding author: \email{oapizio@gmail.com}.}}
\addresses{
\addr{label1}
Departamento de Qu\'{i}mica, Universidad Aut\'{o}noma Metropolitana-Iztapalapa, 
Av. San Rafael Atlixco 186, Col. Vicentina, 09340, CDMX, M\'{e}xico,
\addr{label2}Instituto de de Qu\'{i}mica, Universidad Nacional Aut\'{o}noma de M\'{e}xico,
Circuito Exterior, 04510, Cd. Mx., M\'{e}xico}

\Keywords{molecular dynamics simulation, water-methanol mixtures, partial
molar volumes, excess enthalpy, self-diffusion coefiicients}

\date{Received January 17, 2025, in final form February 10, 2025}

\begin{document}

\maketitle

%%%%%%%%%%%%%%%%%
\begin{abstract}
We report the temperature, pressure and composition dependence of some basic
properties of model liquid water-methanol mixtures.  
For this purpose the isobaric-isothermal molecular dynamics 
computer simulations are employed. Our principal focus is on
the united atom non-polarizable UAM-I-EW model for methanol which was recently parametrized
the paper by Garcia-Melgarejo et al. [~J. Mol. Liq., 2021, \textbf{323}, 114576],
combined with the TIP4P/$\varepsilon$ water model. 
In perspective, the methanol model permits a
convenient extension for other monohydric alcohols mixed with water. 
The behavior of density, excess mixing volume and enthalpy are described.
Partial mixing properties are interpreted. 
Besides, we explored the trends of behavior of self-diffusion coefficients of the
species of a mixture. The quality of predictions of the model
is critically evaluated by detailed comparisons with experimental results.
Various results are novel and provide new insights into the behavior of the
mixtures in question at different temperatures and at high pressures.
An improvement of the modelling necessary for further research is discussed.

\printkeywords
%
%
%\pacs 02.70.Ns,61.20.Ja,82.30.Rs,87.15.hp
\end{abstract}

\section{Introduction}

The present work is an extension of our recent investigations
of the properties of water-methanol liquid mixtures by using
molecular dynamics computer simulations~\cite{galicia1,galicia2,mario1,mario2}.
In contrast to those studies focused on the description of composition 
effects at room temperature and at ambient pressure, here we would like to explore the
trends of behavior of various
properties of water-methanol mixtures on temperature and pressure,
besides the composition changes.

There is no need to say that water is essential for life 
and any kind of human activity. It remains to be a challenging research 
subject from the experimental and theoretical point of view.
It exhibits several thermodynamic anomalies that still lack a definite profound explanation.
The properties of water in both bulk phase and in mixtures
are determined to much extent by changes of the hydrogen bonds network.
Coexistence of water with organic matter determines our being.
One of the principal phenomena within this ``interface'' is the hydrophobic
effect. It refers to the extent of correlations  between nonpolar or amphiphilic 
molecules (solutes) in aqueous media~\cite{sun}.
The strength of this effect depends on the intramolecular structure of solutes,
solute-solvent (water) and solvent-solvent (water-water) interactions.
On the other hand, it depends on the chemical composition of the system 
as well as on temperature and pressure.
This kind of phenomena has implications in physical chemistry 
and biology~\cite{grigera,ghosh,hummer}, as well as
it is of crucial importance for academic research and for practical applications.

One of the simplest amphiphilic molecule types  are alcohols, methanol 
between them in particular. Mixtures of alcohols with water have been studied 
in very many aspects by experimental and theoretical methods for a long time.
This research has generated an enormous amount of literature, practically impossible
to cite comprehensively. Most frequently, these systems were studied upon the
changes of composition at room temperature and at ambient pressure.
In spite of undoubtful importance, much less is known about the behavior
of water-alcohol mixtures upon the changes of temperature and pressure,
probably due to experimental difficulties. These issues are
discussed in various 
publications~\cite{oakenfull,pusztai,yamaguchi,yoshida,sato2,kubota1,kubota2}.

Computer simulation techniques offer alternatives for laboratory research
under such conditions~\cite{moghaddam,koga,mancera,ortega,durell}. 
They represent  useful and popular tools
to get profound insights into the microscopic structure, thermodynamic, 
dynamic,  dielectric and interfacial properties of this type of systems.
The most important, initial step of computer simulations methodology is in
the design of an appropriate force field. The intramolecular structure of alcohol species
is frequently considered at a united atom or all-atom level modelling of non-polarizable
molecules. Unfortunately desirable,  more sophisticated, polarizable force fields require much more
expensive calculations. They are not comprehensively tested and much less
frequently used at present.
The appropriateness of the computer simulation predictions for a given model for
a mixture, upon changing temperature, $T$, pressure, $P$, and composition, $X$, i.e.,
variables, should be tested by comparison with experimental data as much as possible.
Several experimental techniques contributed to the understanding of the
behavior of water-alcohol systems upon changes of temperature, pressure and
composition. Namely, the neutron and dynamic light scattering,
nuclear magnetic resonance, dielectric relaxation, vibrational,
and Raman  spectroscopy, calorimetry --- all of them offer results         
that require support from computer simulations.

Profound insights into the properties of pure components of interest
for the present study from computer simulations are available.
Specifically, a comprehensive set of data for non-polarizable 
water models was provided by Vega and Abascal~\cite{vega-pccp}.
It is commonly accepted that the TIP4P/2005 is almost entirely
successful model in a rather wide interval of thermodynamic parameters.
A similar type of strategy of description was applied to
methanol~\cite{salgado}. A substantial improvement of the dielectric constant
for water was reached, however,  by the development of 
the TIP4P/$\varepsilon$ model~\cite{fuentes}.
Concerning a set of monohydric alcohols, the TraPPE data basis~\cite{trappe} has been
frequently used to deal with individual species and aqueous solutions.
Quite recently, this kind of united atom type modelling was critically reconsidered 
and revised to improve the solubility of alcohols in water~\cite{melgarejo}.
As a result, the so-called UAM-I parametrization of alcohols model was
combined with TIP4P/$\varepsilon$ and tested at ambient pressure and 
temperature~\cite{melgarejo}.  This kind of development was the
subject of very recent study from this laboratory~\cite{bermudez} to yield
a rather comprehensive and successful description of mixtures of 
monohydric alcohols with water upon composition.
Obviously, several combinations of water and alcohol models are necessary to
explore  more in detail.  However, from a general perspective of various alcohols modelling,
it seems interesting to investigate water-methanol mixtures upon temperature and pressure
changes, besides composition. It may provide the first step of an ampler
project involving a set of properties of various systems of this kind. 
Afterwards, one can undertake exploration of these mixtures using all-atom
modelling for the sake of comparisons with united atom level and critical evaluation with
experimental trends.

%%%%%%%%%%%%%%%%%%%%%%%%%%%%%%%%%%%%%%%
%%%%%%%%%%%%%%%%%%%%%%%%%%%%%%%%%%%%%%%%%%%%%%%%%%%%%%%%%%%

\section{Models and simulation details}

In this work we are principally interested in the 
united atom type, non-polarizable model 
with three sites, O, H, CH$_3$, parametrized very recently~\cite{melgarejo}.
For water, the TIP4P/$\varepsilon$~\cite{fuentes} model is considered. 
Nevertheless, in some parts of the manuscript we involve the
united atom TraPPE methanol model~\cite{trappe} and the TIP4P/2005 water model.
In general terms,  within this type of modelling, the interaction  potential 
between all atoms and/or groups is assumed 
as a sum of Lennard-Jones (LJ) and Coulomb contributions. 
Lorentz-Berthelot combination rules are used to determine the cross parameters for
the relevant potential well depths and diameters.

Molecular dynamics computer simulations of water-methanol mixtures have been performed in the
isothermal-isobaric (NPT) ensemble at a  given pressure  and temperature values.
We used GROMACS software~\cite{gromacs} version 5.1.2.
The simulation box in each run was cubic, the total number of molecules of both species 
in all cases is fixed at 3000. The composition of the mixture is described by the molar fraction of methanol
molecules, $X_{2}=N_{2}/(N_{1}+N_2)$, where 1 and 2 refer to water and methanol species
throughout this manuscript.
As common, periodic boundary conditions were used.
Temperature and pressure control has been provided by the V-rescale thermostat and Parrinello-Rahman
barostat with $\tau_T$~=~0.5~ps and $\tau_P$ = 2.0 ps, the timestep was 0.002 ps.
The value of $4.5\cdot 10^{-5}$~bar$^{-1}$ was used for the compressibility of mixtures.

The non-bonded interactions were cut-off at 1.4 nm, whereas the long-range electrostatic interactions
were handled by the particle mesh Ewald method implemented in the GROMACS software package  (fourth
order, Fourier spacing equal to 0.12) with the precision $10^{-5}$. 
The van der Waals correction terms to the energy and pressure were used.
In order to maintain the geometry of water molecules and methanol 
intra-molecular bonds rigid, the LINCS algorithm was used.

After preprocessing and equilibration, consecutive simulation runs, 
each for not less than 10 ns,  with
the starting configuration being the last configuration from the previous
run, were performed to obtain trajectories for the data analysis. 
The results for the majority of  properties  were obtained by averaging 
over 7--10 production runs. 
However, the self-diffusion coefficients were evaluated from the
entire trajectory taking the best slope of the mean squared displacement as common.

\section{Results and discussion}

The first principal issue we would like to deal with is to evaluate the accuracy of  
predictions of the model for water-methanol mixture dependent on temperature, pressure 
and composition. This should be done for a set of target properties. The density is 
one of them. Therefore, we would like to begin this section by revisiting the 
dependence of density of individual species, water and methanol, on $T$, $P$ and $X_2$,
and then proceed to the mixture.

\subsection{Density of water and methanol depending on temperature and pressure}

The dependence of water density, $\rho_1$ on $T$ has been the subject of very many reports
using experimental methods and computer simulation approaches, in part because 
it exhibits an anomalous behavior. These studies were comprehensively described 
in~\cite{sengers}. Here, we involve recent measurements at high pressures 
and the relevant discussion from~\cite{fomin}. One of the first simulation studies that 
reported $\rho_{1}(T)$ curves at different pressures for TIP4P/2005 water model 
is from the laboratory of C. Vega, \cite{helena}. These authors explored an ample interval of
temperatures, starting from $T \approx 375$~K  down to deeply supercooled water.
Four pressure values, $P = 1$ bar, 400 bar, 1000 bar and 1500 bar were considered.
However, the simulations were performed for a small system of 256 and 500 molecules
with cutoff of interactions at 0.9~nm, to obtain density and isothermal compressibility. 
Therefore, some of these simulations were redone for a bigger system using TIP4P/2005 model 
and a larger cutoff for inter-particle interaction, as noted above. 
However, the TIP4P/$\varepsilon$ model is of 
our principal interest. Previous comparison of the predictions from 
TIP4P/2005 and TIP4P/$\varepsilon$ was performed in~\cite{fuentes} using 
500  TIP4P/$\varepsilon$  water molecules with $r_{\text{cut}}=0.95$ nm. Actually, we would
like to confirm some of the conclusions from~\cite{fuentes}.

\begin{figure}[!t]
\begin{center}
\includegraphics[width=6.0cm,clip]{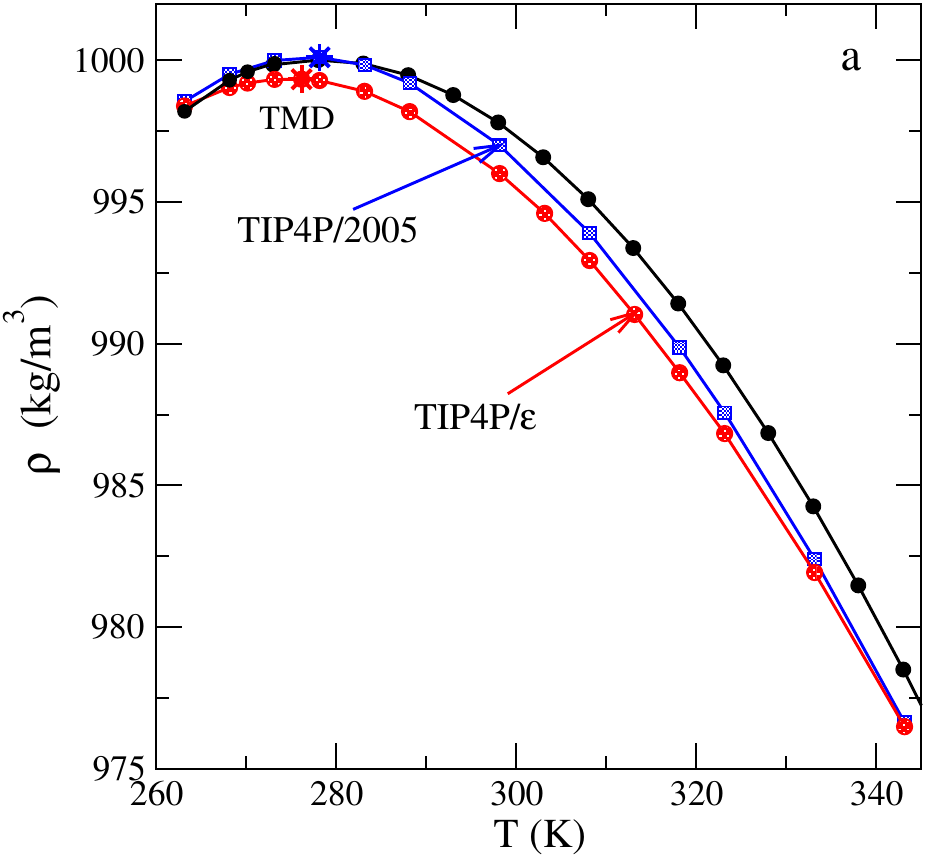}
\includegraphics[width=6.0cm,clip]{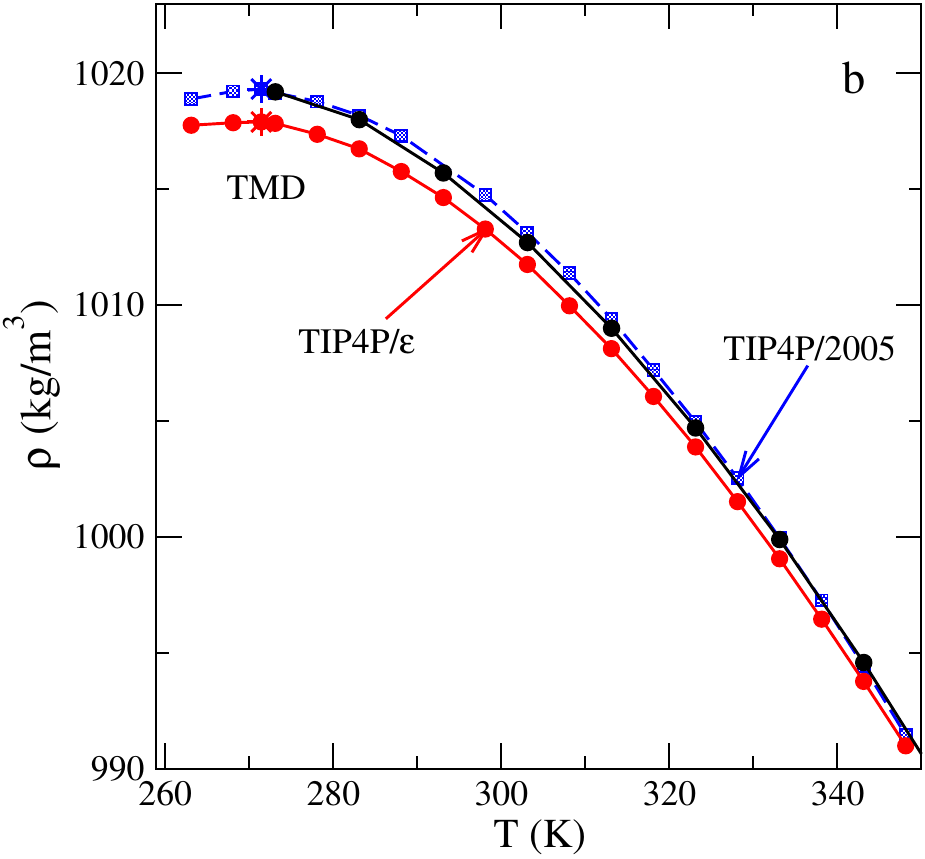}
\end{center}
\caption{(Colour online) 
Water density depending on temperature
at pressure 1 bar (panel a) and at 400 bar (panel b).
Black circles denote experimental data. In panel a 
they are taken from~\cite{fomin}, three experimental points at the 
lowest temperatures are from \cite{hare} for supercooled water. 
In panel b they come from~\cite{milero}. 
The blue squares and red circles are our simulation results for
TIP4P/2005 and TIP4P/$\varepsilon$ models, respectively.
Stars denote the temperature of maximum density (TMD).
}
%\protect
\label{fig1}
\end{figure}

\begin{figure}[!t]
\begin{center}
\includegraphics[width=6.0cm,clip]{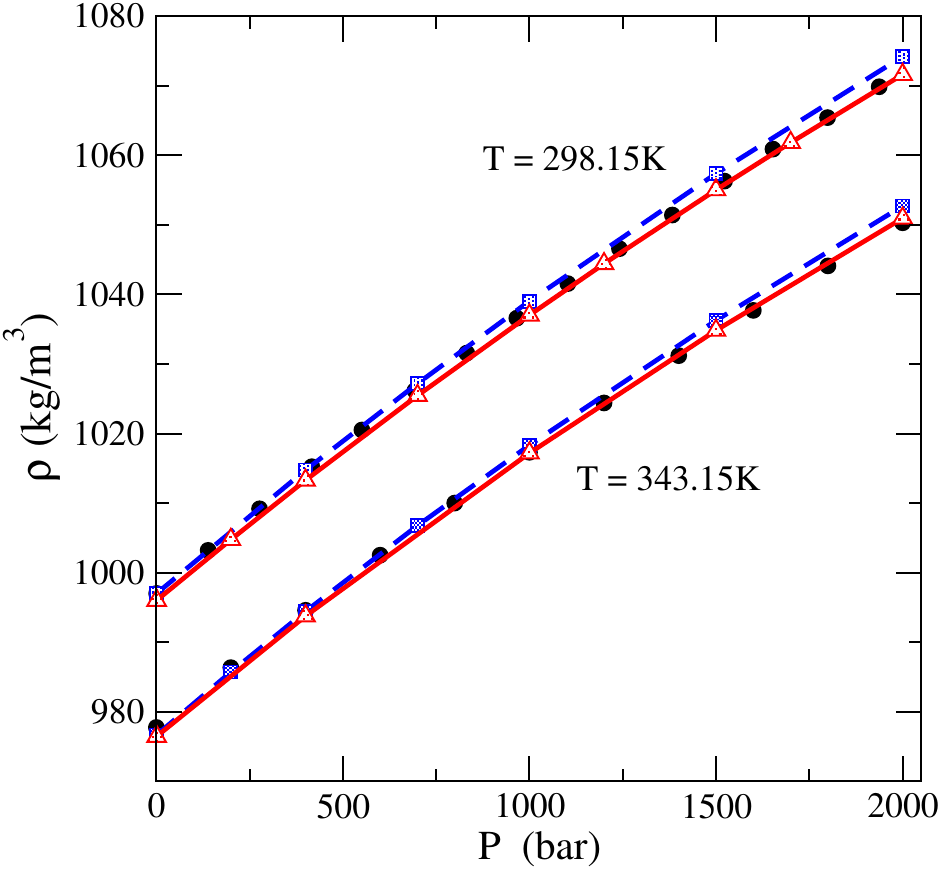}
\end{center}
\caption{(Colour online)
The dependence of water density on pressure at 298.15~K and at 343.15~K.
The experimental data (black circles) are from~\cite{kubota1} (298.15K) and
NIST Chemistry Webbook (343.15~K)~\cite{nist}, respectively.
The nomenclature of lines and symbols as in figure~\ref{fig1} (blue --- TIP4P/2005, red --- TIP4P/$\varepsilon$).
}
\label{fig2}
\end{figure}

Our calculations confirm three issues. Both water models, TIP4P/2005 and TIP4P/$\varepsilon$
reproduce the experimental curve perfectly well. They underestimate water density at temperatures 
above 298.15~K very slightly (\ref{fig1}a). The temperature of maximum density (TMD) is reproduced
at $T = 276.15$~K for TIP4P/2005 model, the TIP4P/$\varepsilon$ predicts it at almost
the same temperature, 275.15~K. Moreover, both models correctly predict the shift of TMD
to a lower temperature upon increasing pressure from 1 bar to 400 bar (\ref{fig1}b).
Apparently, the TIP4P/2005 is a bit superior to TIP4P/$\varepsilon$ in this interval of
parameters. On the other hand, if one considers the pressure dependence of density at $T = 298.15$~K
and $T = 343.15$~K, then the TIP4P/$\varepsilon$ model exhibits a better performance at rather
high values of pressure (\ref{fig2}2).

\begin{figure}[!t]
\begin{center}
\includegraphics[width=5.5cm,clip]{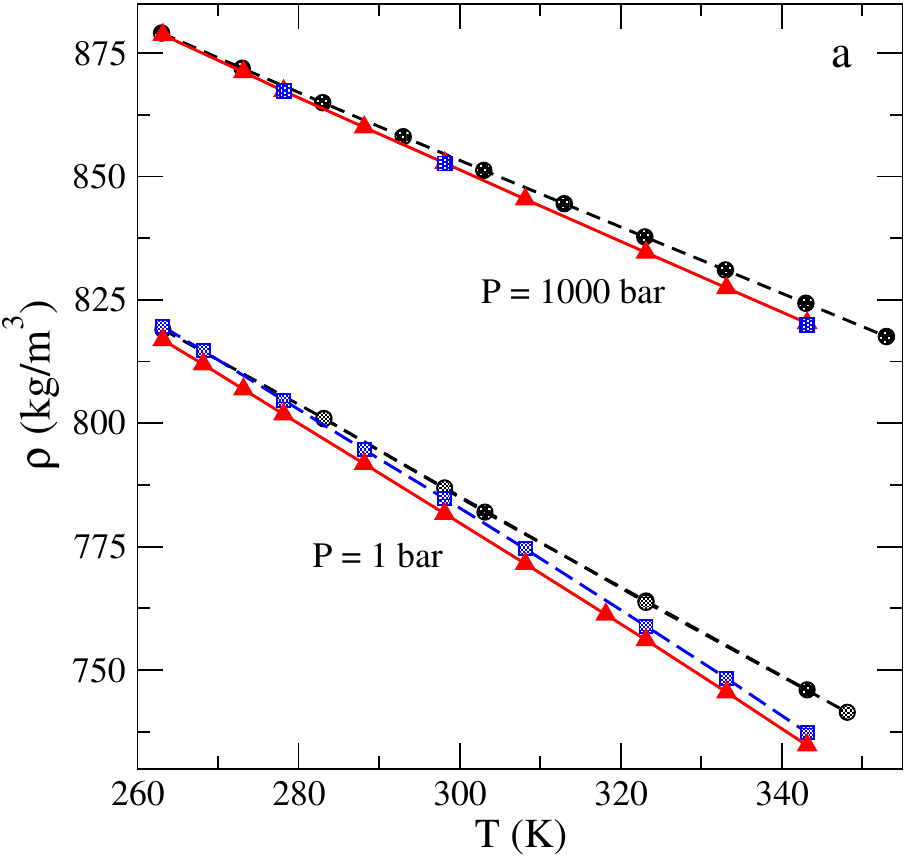}
\includegraphics[width=5.5cm,clip]{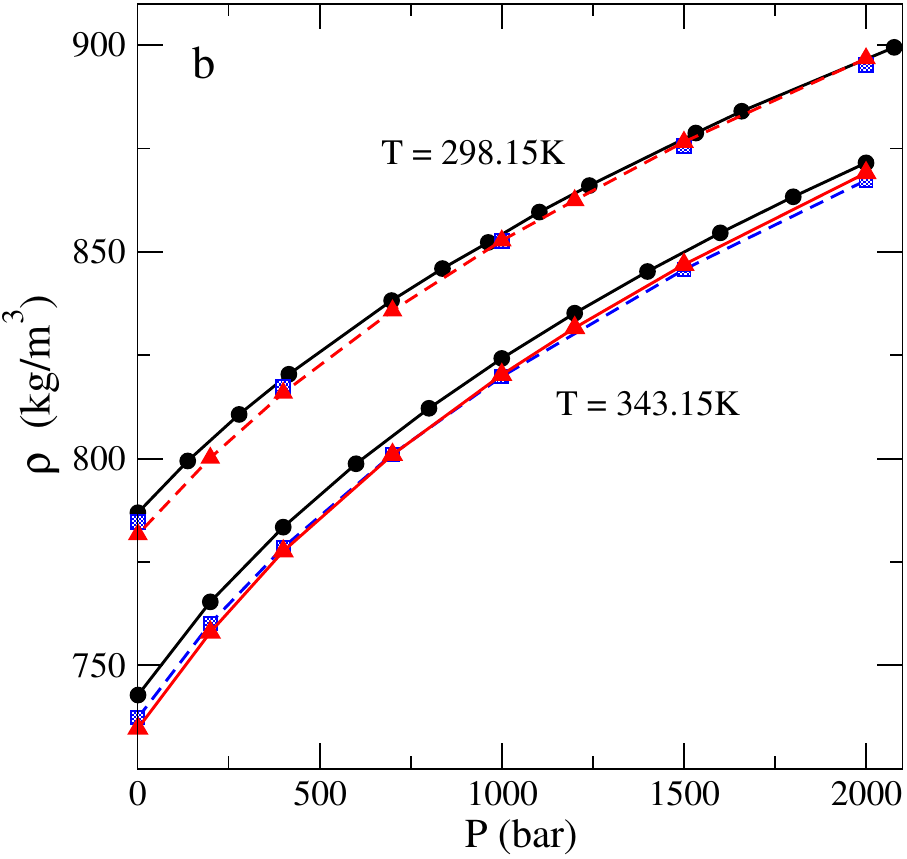}
\end{center}
\caption{(Colour online) Panel a: Methanol density depending on temperature
at pressure 1 bar,  and at 1000 bar.
The simulation results are for  UAM-I united atom methanol model (red triangles)
and TraPPE model (blue squares). The experimental data 
(black circles) at 1 bar are from Engineering toolbox  
and  Kubota et al.~\cite{kubota1}, while at 1000 bar they are from~\cite{nist}. 
Panel b: Methanol density depending on pressure at temperature 298.15~K and at 343.15~K.
The experimental data are from~\cite{kubota1} (298.15K) and~\cite{nist} (343.15). 
Other notations as in panel a.
}
\label{fig3}
\end{figure}

Next, we performed simulations of pure methanol using the UAM-I and TraPPE united atom models.
Both models describe the temperature dependence of density quite well at normal pressure,
1 bar, and at much higher pressure, 1000 bar (figure~\ref{fig3}a). Nevertheless, the deviation 
of simulation results from experimental data grows smoothly upon increasing temperature.
At $P = 1$ bar, the TraPPE model is a bit better than the UAM-I. At a high pressure, however,
the results of two models flow together. In panel b of this figure, figure~\ref{fig3}b, the methanol 
density as a function of pressure is shown. Both methanol models  reproduce 
$\rho (P)$  dependence very well. Deviation from the experimental data is rather small and 
decreases upon increasing pressure. In summary, we have confidence that two constituents
of the mixtures in question, in their pure state, are well described.

\subsection{Composition dependence of density of methanol-water mixtures at different temperatures and pressures}

As we have mentioned in the introductory section, there have been several
experimental reports concerning the density of water-methanol  mixtures 
upon changing composition. We used experimental data at room temperature $T= 298.15$~K, and at
atmospheric pressure~\cite{mikhail}. At  higher pressures, the experimental data 
are scarce. We have found solely the report by Kubota et al.\cite{kubota1}.

\begin{figure}[!t]
\begin{center}
\includegraphics[width=6.5cm,clip]{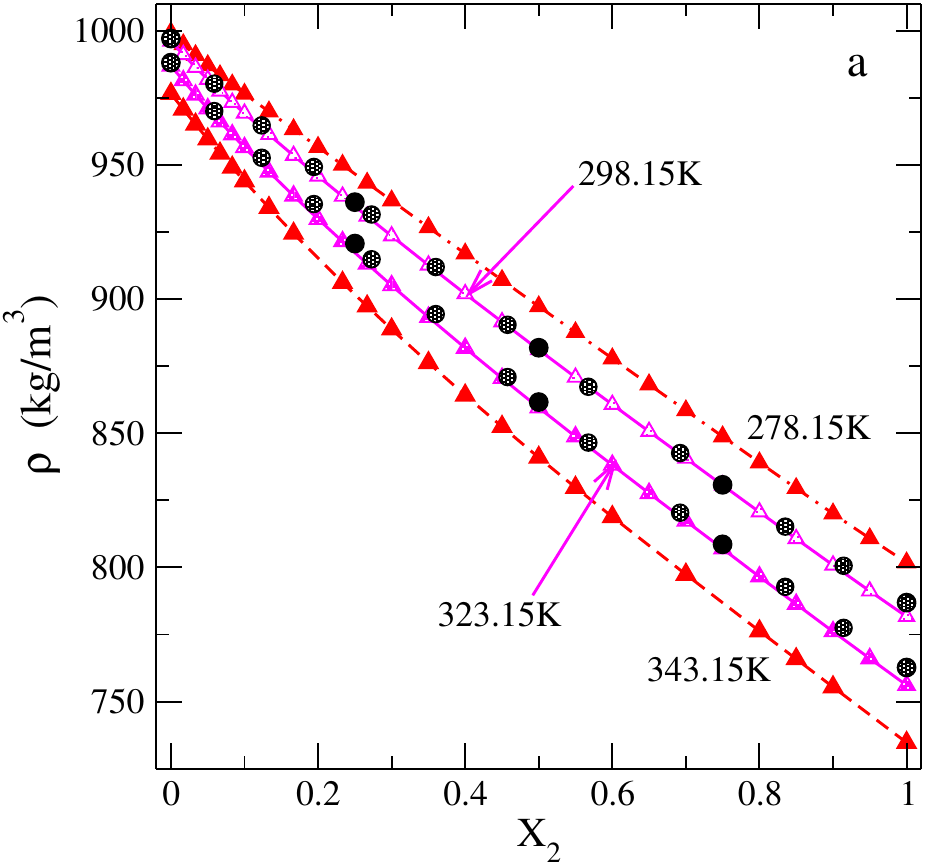}
\includegraphics[width=6.5cm,clip]{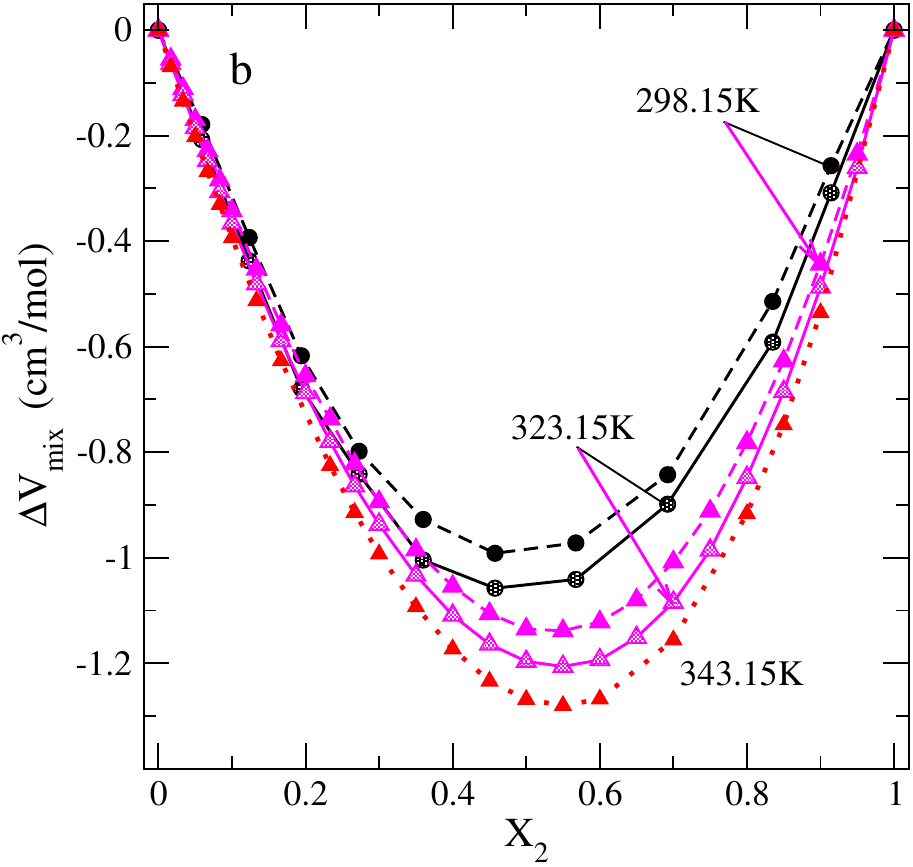}
\end{center}
\caption{(Colour online) Panel a: Composition dependence of density of water-methanol
mixtures at different temperatures, at a fixed pressure, $P = 1$ bar. 
The experimental data (black circles), at $T=298.15$~K and $323.15$~K, 
are from~\cite{kubota1,mikhail}, respectively. 
The simulation results (red and magenta triangles) are 
for TIP4P/$\varepsilon$ --- UAM-I model.
Panel b:  The excess mixing volume depending on composition at different temperatures
at a fixed pressure, $P = 1$ bar. Experimental data are taken from~\cite{mikhail}.
}
\label{fig4}
\end{figure}

We learn from figure~\ref{fig4}a that the $\rho (X_2)$ dependencies at different temperatures
behave similarly. They are almost linear curves descending from pure water to pure
methanol values. From the limited comparison with experimental data, one can conclude
that the behavior of density is reasonably correct. However, small inaccuracy of simulation
models can be seen for methanol-rich mixtures.

It is important to correctly capture the deviation from ideality 
of a given property. This kind of insights follows, for example, from the excess mixing volume. 
The excess mixing volume is defined as
follows, $\Delta V_{\text{mix}} = V_{\text{mix}} - X_{2} V_{2} - (1-X_{2}) V_1$,
where $V_{\text{mix}}$, $V_{2}$ and $V_1$ refer to the molar volume of the mixture and
of the individual components, methanol and water, respectively.
Experimental data show that $\Delta V_{\text{mix}}$ is negative and exhibits
a minimum at $X_{2} \approx 0.45$, figure~\ref{fig4}b.
The simulation results show qualitatively similar trends of behavior.
A comparison between the experiment and simulations
with TIP4P/$\varepsilon$--UAM-I model shows that the model describes $\Delta V_{\text{mix}}$
for water-rich mixtures very well. At a higher alcohol content, $X_2 > 0.2$, the model
overestimates the values for $\Delta V_{\text{mix}}$. The minimum of the excess mixing volume
from simulations is at $X_{2} \approx 0.55$ at different temperatures under study.
The values for the $\Delta V_{\text{mix}}$ at a fixed value for $X_2$ slightly increase 
in magnitude with increasing temperature, presumably due to the break of hydrogen bonds.
The simulation model appropriately reproduces $\Delta V_{\text{mix}} (X_2)$ and its temperature
changes. Thus, an overall performance of the model can be termed as satisfactory
with qualitative accuracy. It would be fair to note that the TIP4P/2005--TraPPE
model, studied by us previously, provides a better description of $\Delta V_{\text{mix}} (X_2)$
at a single studied temperature,  $T = 298.15$~K  and at $P = 1$~bar, cf. figure~2a of~\cite{mario1}.

\begin{figure}[!t]
\begin{center}
\includegraphics[width=6.5cm,clip]{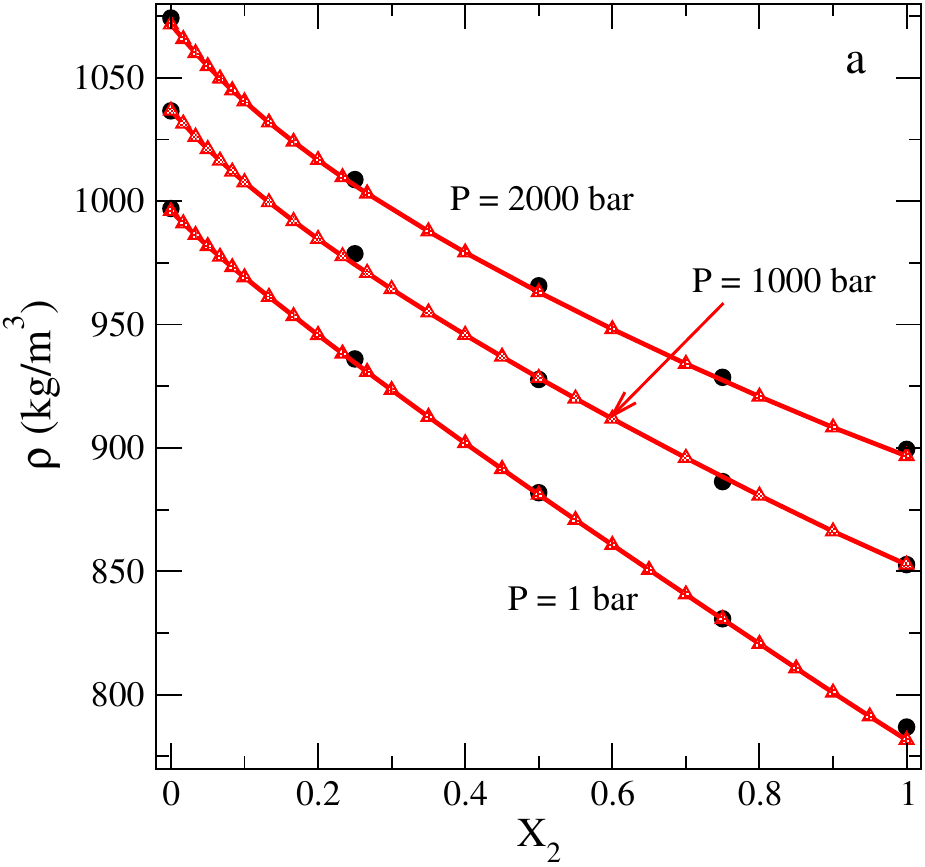}
\includegraphics[width=6.5cm,clip]{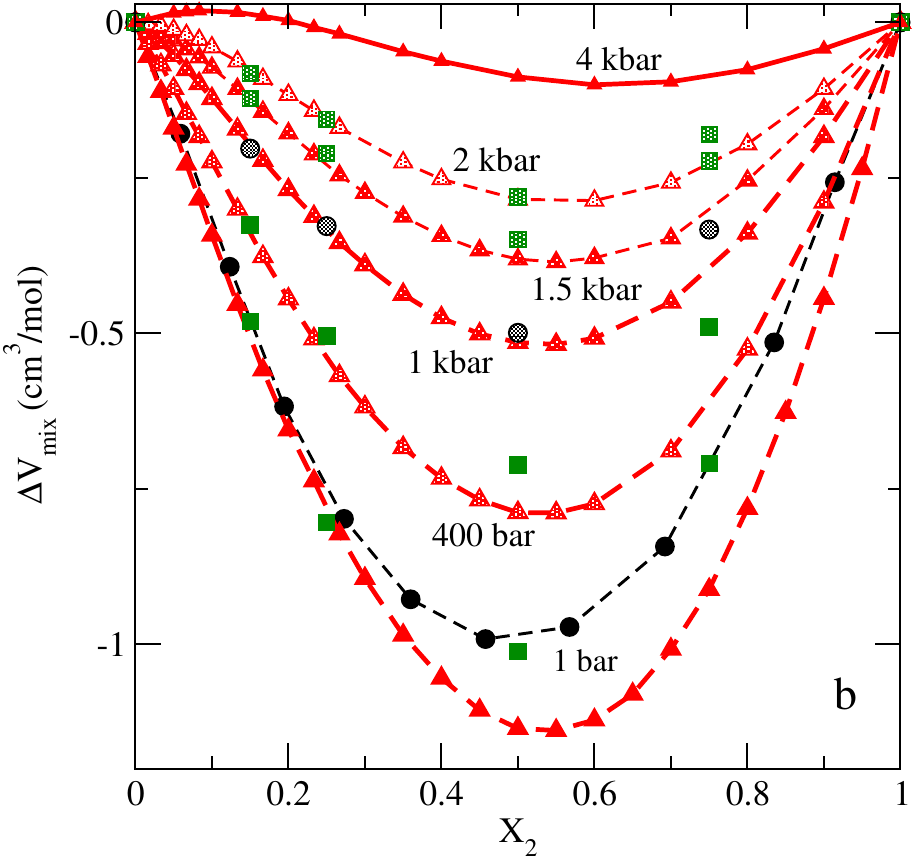}
\end{center}
\caption{Panel a: Composition dependence of density of water-methanol
mixtures at different pressures, $P = 1$ bar, 1000 bar, and 2000 bar 
at a fixed temperature 298.15~K (panel a). Panel b: Composition dependence
of excess mixing volume at different pressures indicated in the
figure at 298.15~K. Experimental data in panel b are from~\cite{mikhail}
(black circles) and from~\cite{kubota2} (green squares).
}
\label{fig5}
\end{figure}

Now, we would like to perform similar analyses, as in figure~\ref{fig4}, exploring the behavior
of $\rho (X_2)$ at different pressures and at a fixed value of temperature, $T = 298.15$~K.
The simulation results for the TIP4P/$\varepsilon$--UAM-I model are shown in figure~\ref{fig5}a.

It follows that the density of the mixtures depending on composition at different pressures
is described quite satisfactorily. Therefore, the excess mixing volume depending on composition
at different pressures is satisfactory as well. Moreover, the agreement between
computer simulation results and experimental data becomes better with increasing pressure.
The location of the minimum of $\Delta V_{\text{mix}} (X_2)$ is also rather well captured
by the model considered at high pressures. However, certain discrepancy is observed for
methanol-rich mixtures.

\begin{figure}[!t]
\begin{center}
\includegraphics[width=5.5cm,clip]{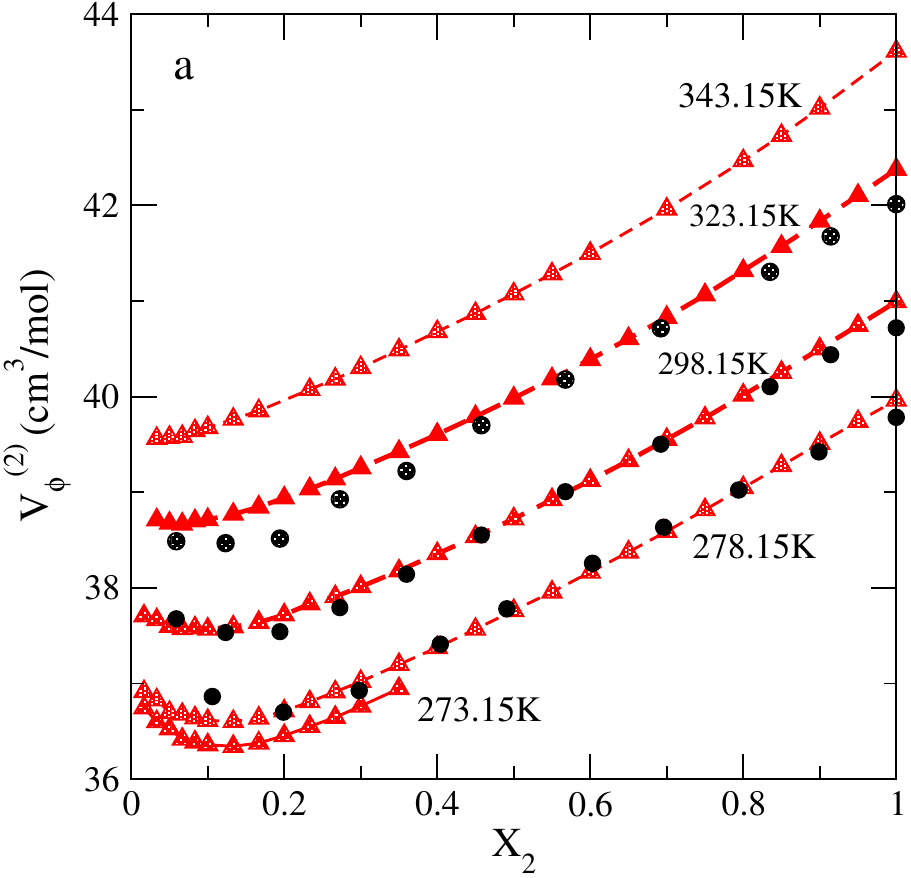}
\includegraphics[width=5.5cm,clip]{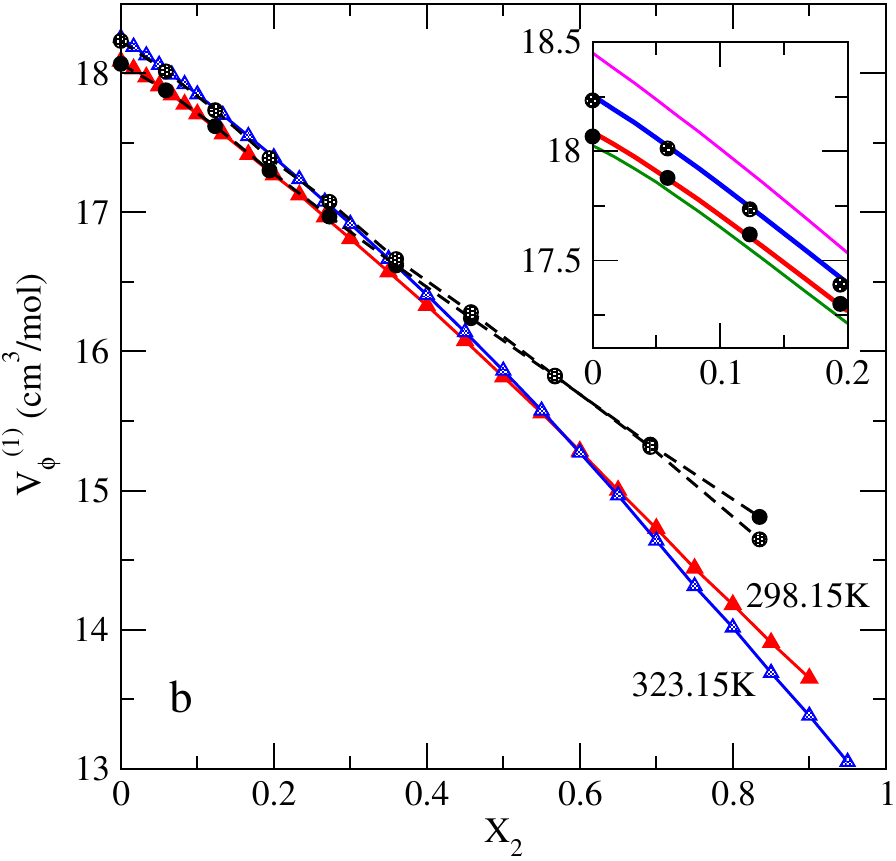} \\
\includegraphics[width=5.5cm,clip]{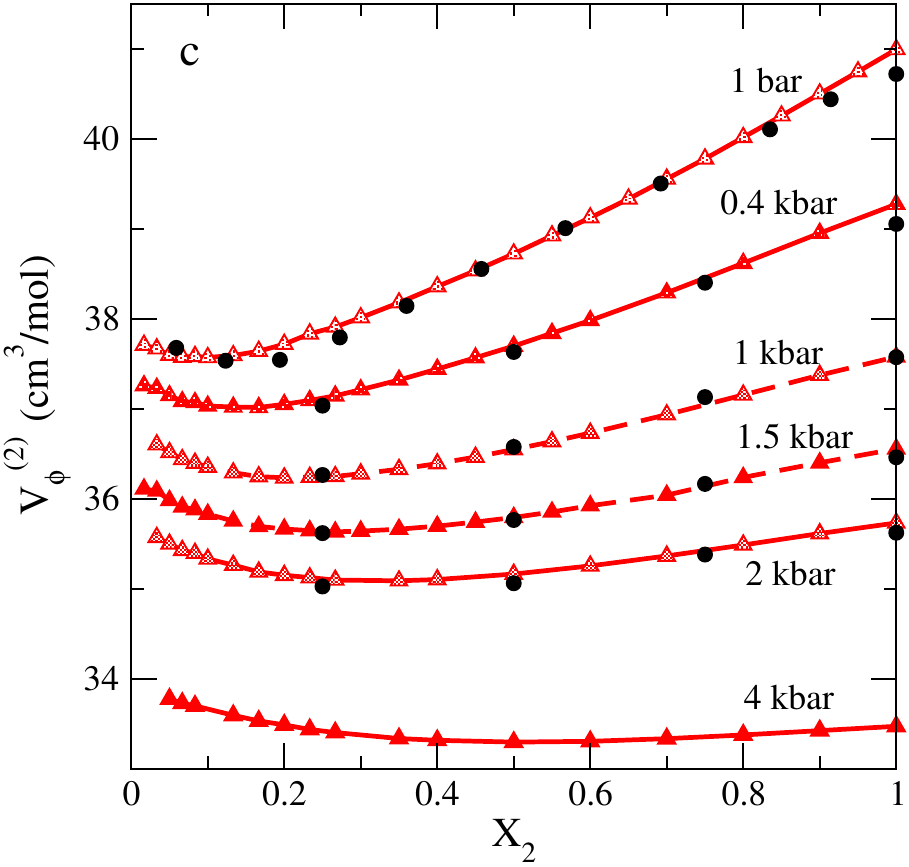}
\includegraphics[width=5.5cm,clip]{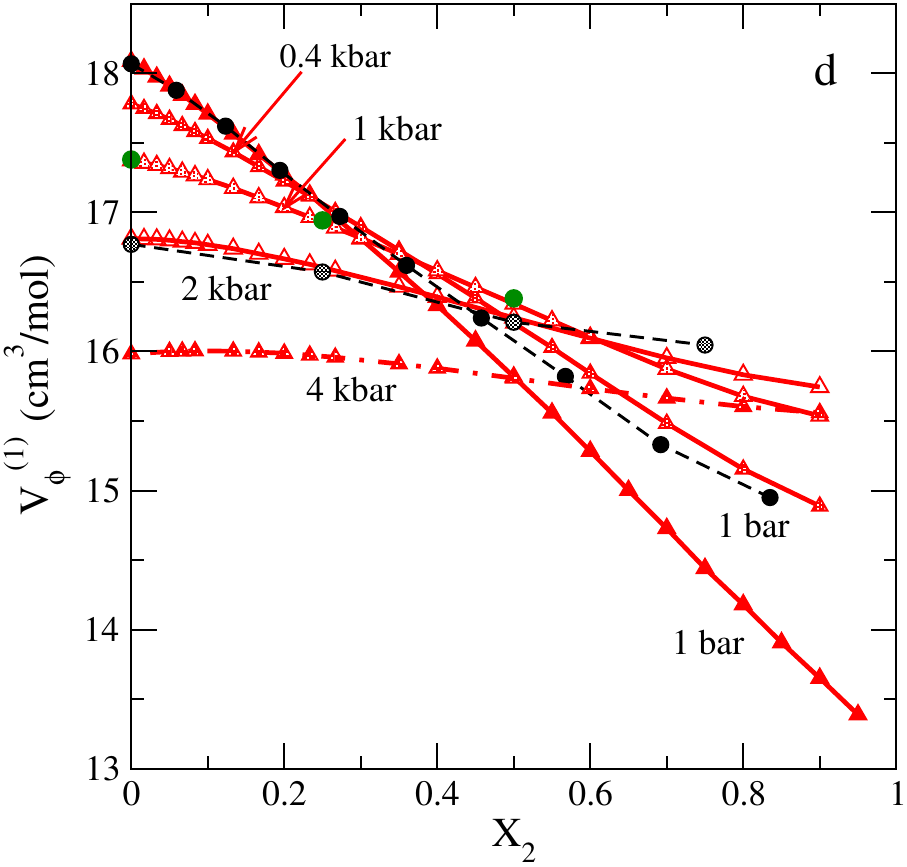}
\end{center}
\caption{(Colour online)
Panels a and b: A comparison of the composition dependence of the
apparent molar volumes of methanol and water species from simulations,
with the experimental data at different temperatures. The experimental
data (black circles) are from~\cite{mikhail} at 298.15~K and 323.15~K,
and from Easteal et al.~\cite{easteal} at 278.15~K. The curves in the inset to panel b
are for $T=278.15$~K (green),
$T=298.15$~K (red), $T=323.15$~K (blue) and $T=343.15$~K (magenta).
Panels c and d illustrate the behavior of the apparent molar volumes at different
pressures. The temperature is fixed at 298.15~K.
}
\label{fig6}
\end{figure}

In order to discern the contributions of each species into the excess molar volume
and to obtain deeper insights into the geometric aspects of mixing depending on composition,
both from experiments and simulations, one can resort to the notion of
the apparent molar volume of the species rather than the excess molar volumes.
The apparent molar volume
for each species according to the definition is~\cite{torres}:
$V_{\phi}^{(1)}= V_1 + \Delta V_{\text{mix}}/(1-X_{2})$
and  $V_{\phi}^{(2)}= V_{2} + \Delta V_{\text{mix}}/X_{2}$.
We elaborated the experimental density data from~\cite{mikhail}
and the results from our simulations to construct the plots shown in 
panels a, b, c and d of figure~\ref{fig6}.

The TIP4P/$\varepsilon$--UAM-I  model provides a quite accurate description of the
composition behavior for $V_{\phi}^{(2)}$ in water-rich mixtures 
and in the entire composition range at a fixed pressure, $P =1$ bar (figure~\ref{fig6}a).
The minimum of  $V_{\phi}^{(2)}$ is predicted at a slightly
lower methanol concentration, $X_{2} \approx 0.1$ ($T = 298.15$~K), 
in comparison to the experimental result, $X_{2} \approx 0.13$.
The minimum becomes more pronounced if the temperature decreases from 298.15~K to 
273.15~K. On the other hand, the minimum is hardly visible at 323.15~K and is absent
at a higher temperature, 343.15~K. Moreover, the location of the minimum on composition axis
shifts to lower values of $X_2$ upon increasing temperature. 
The level of accuracy of the present modelling is similar to the predictions 
of temperature trends for $V_{\phi}^{(2)}$ from TIP4P/2005--TraPPE model
in figure~3a of~\cite{mario2}.

Similar trends of behavior for $V_{\phi}^{(2)}$ are observed  at different pressures, 
but at a fixed value of temperature, $T = 298.15$~K  (figure~\ref{fig6}c). The minimum of $V_{\phi}^{(2)}$
shifts to higher values of $X_2$ upon increasing pressure. The experimental data
definitely confirm the
existence of the minimum at 1 bar only. The lack of experimental data 
at higher pressures precludes to
make conclusions even at a moderate pressure, $P  = 400$~bar. 
We performed computer  simulations up to $P = 4$~kbar to establish the existence of
the limiting pressure at which the minimum ceases to exist. Still, the minimum
of $V_{\phi}^{(2)}$ is observed at 4 kbar.

\begin{figure}[!t]
\begin{center}
\includegraphics[width=6.5cm,clip]{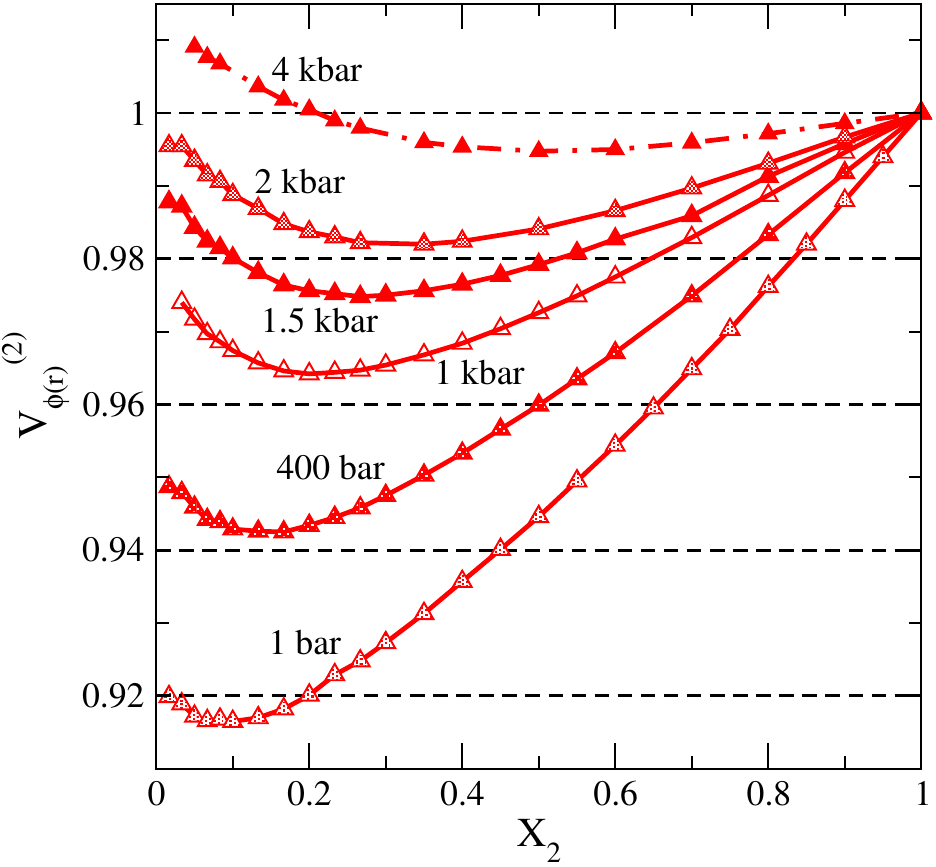}
\end{center}
\caption{(Colour online)
Composition dependence of the reduced
apparent molar volume of methanol 
in water-methanol mixtures from simulations of TIP4P/$\varepsilon$--UAM-I  model
at different pressures. 
The temperature is fixed at 298.15~K.
}
\label{fig7}
\end{figure}

The behavior of $V_{\phi}^{(1)}$ depending on temperature and pressure in panels b and d
of figure~\ref{fig6} is less illuminating. It is necessary to mention, however, that
the simulation results agree well with the experimental data for mixtures
with dominating water content only. By contrast, the model predictions deviate
from the experimental results for $V_{\phi}^{(1)}$ in methanol-rich solutions. The change of
inclination of $V_{\phi}^{(1)}$ on $X_2$ upon increasing temperature
(figure~\ref{fig6}b) or with increasing pressure (figure~\ref{fig6}d), is well observed.
However, interpretation of these trends would require a description of changes of
the microscopic structure of water subsystem. Therefore, these observations 
will be revisited in the following publication on the subject.

In order to obtain a summarizing insight into the behavior of $V_{\phi}^{(2)}(X_2)$
in the entire pressure interval, we have reconsidered the simulation data from figure~\ref{fig6}c
and constructed a similar plot by using a reduced apparent volume of methanol
species, $V_{\phi (r)}^{(2)}(X_2)$ (figure~\ref{fig7}). It is defined as follows, 
$V_{\phi (r)}^{(2)}(X_2) = V_{\phi}^{(2)}(X_2)/V_{\phi}^{(2)}(X_2=1)$.
From this kind of representation, one can see that at a high pressure, close to 4~kbar,
the reduced apparent volume of alcohol species exhibits ``crossover'' from $V_{\phi (r)}^{(2)}(X_2) < 1$
behavior to $V_{\phi (r)}^{(2)}(X_2) > 1$, in the water-rich mixtures. In other words,
expansion (with respect to the molar volume of pure methanol) rather than contraction 
is observed. It is difficult to interpret this behavior on its own, 
without referring to other properties.

In this aspect we would like to recall that the composition behavior of the apparent 
molar volume of alcohol species
in water-rich mixtures can be related to the experimental results for abnormal
intensity of scattered light~\cite{kojima,brill1,brill2}.
Interestingly, it has been found that the Brillouin scattering behavior for 
water-methanol mixtures changes at $\approx$ 4~kbar according to the 
experiments discussed in~\cite{brill1}. These observations support
the existence of a specific high pressure value leading to the peculiarities of geometric
arrangement of species in water-methanol mixtures. Our simulation data
confirm this fact in terms of $V_{\phi (r)}^{(2)}(X_2)$.
In general terms, the simulation findings concerning the peculiarities of
$V_{\phi}^{(2)}(X_2)$ behavior depending on $T$ and $P$ with a corresponding minimum, can urge the 
experimental work along this line of research.

%%%%%%%%%%%%%%%%%%%%%%%%%%%%%%%%%%%%%%%%%%%%%%%%%%%

\subsection{Energetic aspects of mixing of ethanol and water molecules}

\begin{figure}[!t]
\begin{center}
\includegraphics[width=6.5cm,clip]{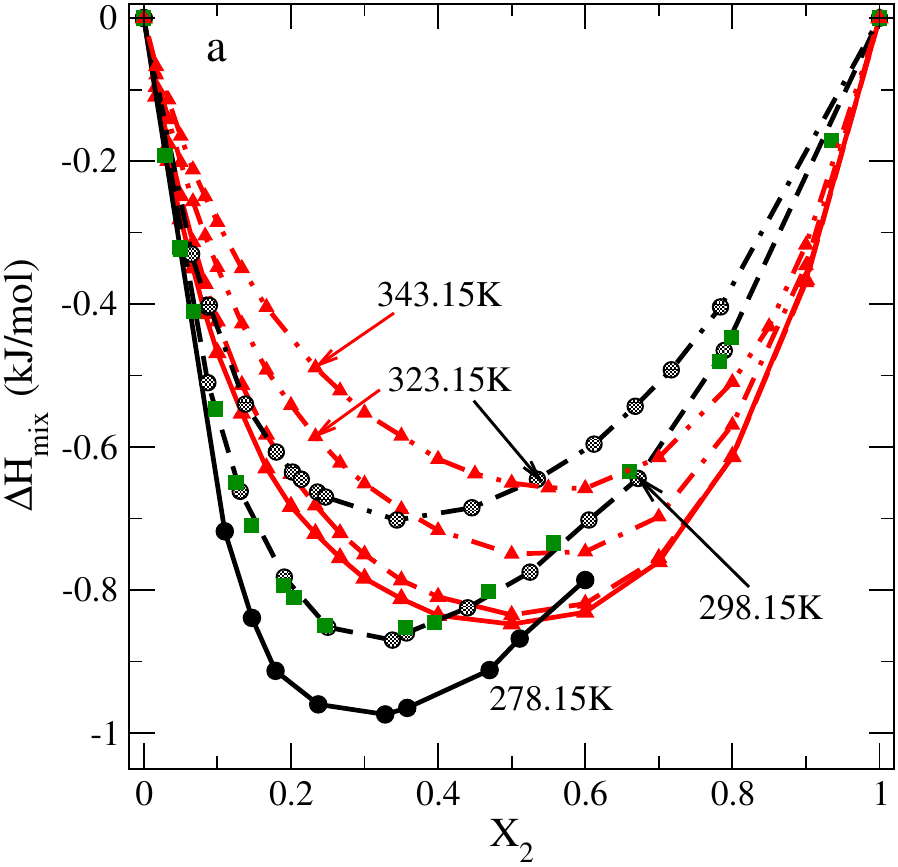}
\includegraphics[width=6.5cm,clip]{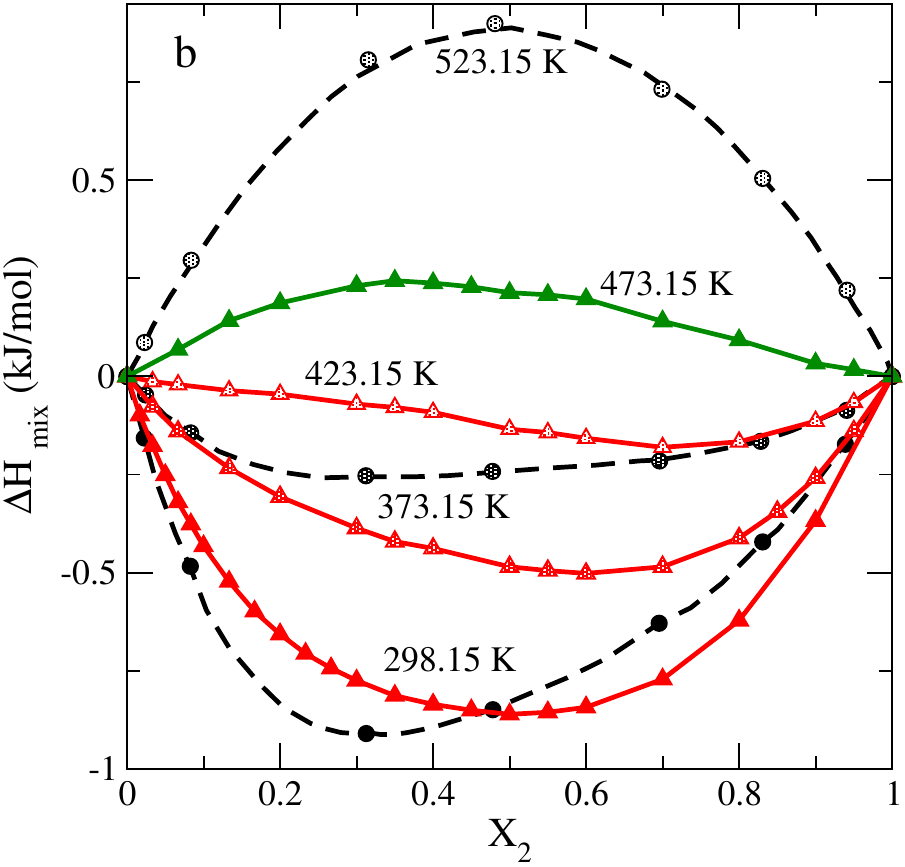}\\
\includegraphics[width=6.5cm,clip]{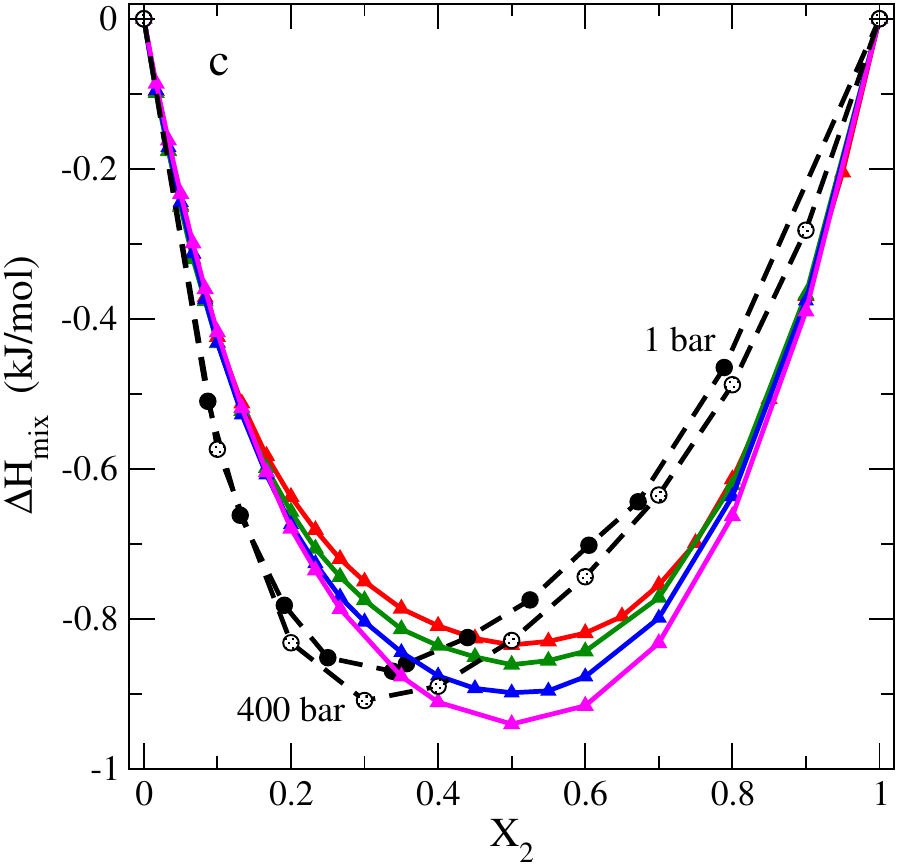}
\end{center}
\caption{(Colour online) Panel a: A comparison of the behavior of the excess mixing enthalpy depending
on temperature from simulations of TIP4P/$\varepsilon$--UAM-I model (red triangles)  
and experimental data (black circles at 323.15~K, 298.15~K, and 278.15~K) from~\cite{iwona}. 
Another set of experimental points (green squares) for 298.15~K are from~\cite{lama}. 
The pressure is fixed at 1 bar.
Panel b: Temperature changes of $\Delta H_{\text{mix}}(X_2)$ from simulations, experimental
data (black circles and polynomial fitting curves) were reproduced
from figure~1c of~\cite{simonson} at 400 bar.
Panel c: Pressure changes of $\Delta H_{\text{mix}}(X_2)$ at a fixed
temperature 298.15~K. The simulation data (triangles) are
for the TIP4P/$\varepsilon$--UAM-I model at 1, 400, 1000 and 2000 bar 
from top to bottom (red, green,
blue and magenta lines), respectively. 
The experimental data (circles) are from~\cite{iwona}.
}
\label{fig8}
\end{figure}

Energetic manifestation of mixing trends is commonly discussed in terms of
the excess mixing enthalpy, $\Delta H_{\text{mix}}$. It is defined similar to the excess mixing 
volume above. We used the experimental results from~\cite{iwona,lama,simonson} and our simulation data to explore $\Delta H_{\text{mix}}$ 
upon composition of water-methanol mixtures. The results are given in figure~\ref{fig8}.
It can be seen that $\Delta H_{\text{mix}}$ values coming from the simulation model 
are underestimated in comparison with experimental data at a 
low temperature, $T = 278.15$~K (panel a).
However, this trend reverts at a higher temperature, $T = 323.15$~K.
At this temperature, the model a bit overestimates the magnitude of
values for $\Delta H_{\text{mix}}$ at compositions around minimum. Moreover, the minimum
of $\Delta H_{\text{mix}}$ from simulation data is located at a higher $X_2$ compared to
the experimental data. In summary, the model correctly predicts the temperature trends.
The effect of attractive inter-molecular interactions decreases
with increasing temperature. In other words, $\Delta H_{\text{mix}}$ values decrease
in absolute values.
The shape of $\Delta H_{\text{mix}} (X_2)$ is reproduced qualitatively correctly as well.
However, the quantitative agreement of simulation results and experimental points
is not reached at ambient pressure, $P = 1$~bar (panel a).

We were guided by the experimental data from~\cite{simonson} to perform
simulations and construct a set of curves in panel b of figure~\ref{fig8}. These results
illustrate the changes of balance between attractive and repulsive forces in the
system upon the changes of temperature at a fixed pressure, $P = 400$~bar.
According to the simulation predictions, the attractive forces still dominate at 
$T = 423.15$~K to yield negative values for $\Delta H_{\text{mix}}$ in the entire composition
interval. However, at a higher temperature, $T = 473.15$~K, $\Delta H_{\text{mix}}$ is
positive for all $X_2$. The experimental data indicate that such kind of change
from exothermic to endothermic mixing
occurs within the temperature interval between $373.15$~K and 523.15~K. Unfortunately, intermediate
experimental data are missing. On the other hand, it is difficult
to establish the trends of behavior of the extremum, $\partial \Delta H_{\text{mix}}/\partial X_2 =0 $ depending
on temperature. The simulation predictions imply that the extremum composition
increases with an increasing temperature in the temperature interval of 
negative values of $\Delta H_{\text{mix}}$. Apparently, the ``crossover'' occurs in the
methanol-rich mixtures. However, if the repulsive interactions become dominating
in the system at $T = 473.15$~K, the extremum composition corresponds to $X_2 \approx 0.4$,
i.e. for mixtures of the type $2 \text{MeOH} - 3 \text{H}_2\text{O}$. This kind of change, despite
the experimental data being scarce, deserves a more extensive study using computer
simulations at different values of pressure.

Concerning the trends of behavior of $\Delta H_{\text{mix}} (X_2)$ at different pressures
and at a fixed temperature, $T = 298.15$~K, one should note a reasonable agreement between
simulations and experiment. It can be termed as satisfactory at a qualitative level.
Namely, the dependence on pressure is correct within this
pressure interval, the magnitude of $\Delta H_{\text{mix}} (X_2)$
values increases upon increasing pressure from 1 bar to 400 bar. Indeed, the 
simulations predict this kind of behavior up to 2000 bar. However, in quantitative
terms, the growth of magnitude of $\Delta H_{\text{mix}}$ at minimum is modest.
It mirrors the observations shown in the previous figure~\ref{fig6}c for the minimum of the apparent
molar volume of alcohol species. Probably, one should explore even higher values
of pressure to intuitively capture a possible reversal of the $\Delta H_{\text{mix}}$ behavior.

In order to elucidate the reasons of the discrepancy of modelling and
experimental predictions for $\Delta H_{\text{mix}}(X_2)$, it is worth to resort to the 
partial excess molar enthalpies.
They follow from the excess mixing enthalpy, $\Delta H_{\text{mix}}$ according to the
definition~\cite{torres},
\begin{equation}
h_{\text{ex}}^{(1)} =\Delta H_{\text{mix}}+X_{2}\left(\frac{\partial \Delta H_{\text{mix}}}{\partial X_1}\right)\Big{\vert}_{P,T}\,,
\end{equation}
\begin{equation}
h_{\text{ex}}^{(2)} =\Delta H_{\text{mix}} -X_1\left(\frac{\partial \Delta H_{\text{mix}}}{\partial X_1}\right)\Big{\vert}_{P,T}\,,
\end{equation}
where, $X_1 = 1- X_{2}$.

\begin{figure}[!t]
\begin{center}
\includegraphics[width=6.5cm,clip]{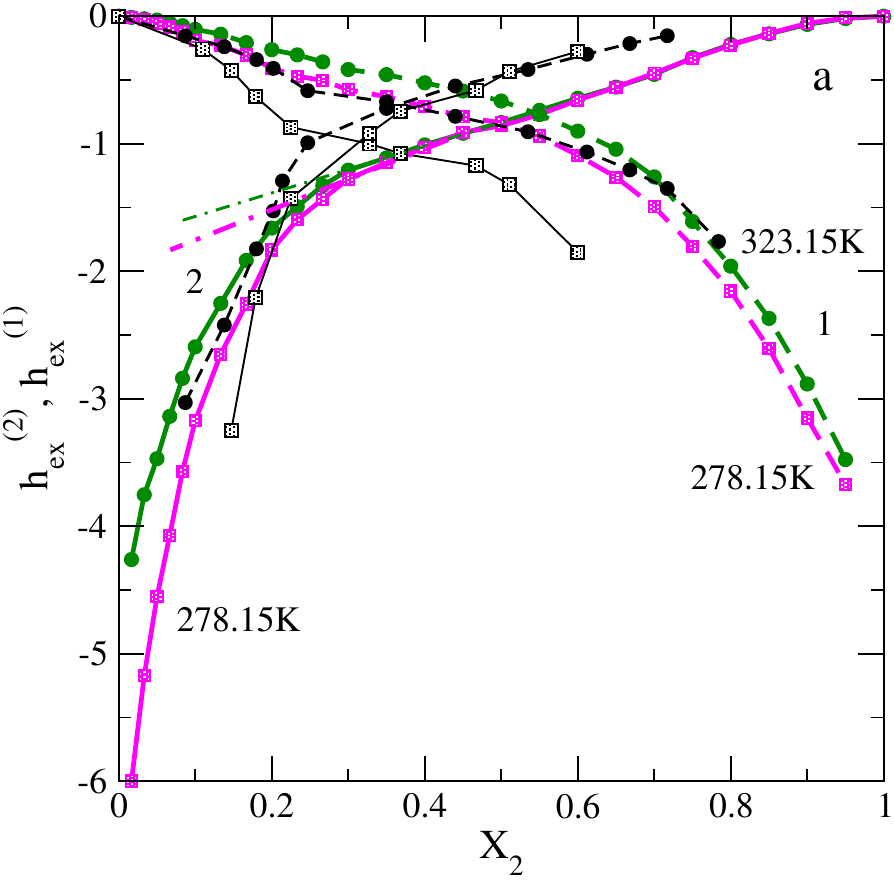}
\includegraphics[width=6.5cm,clip]{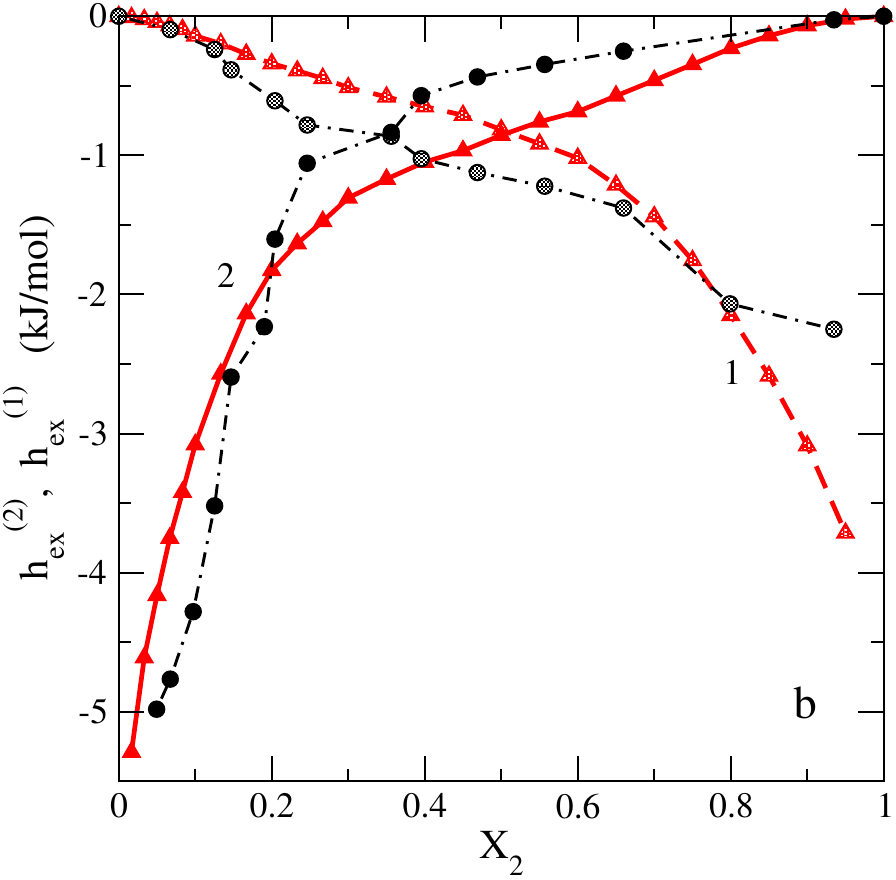}\\
\includegraphics[width=6.5cm,clip]{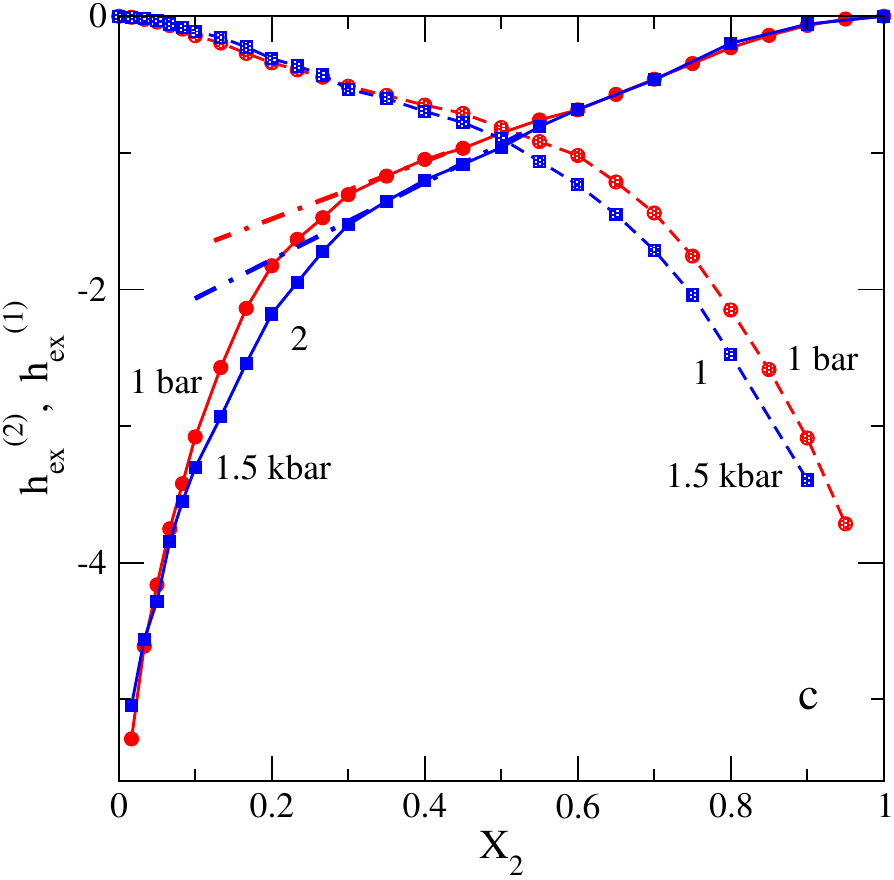}
\end{center}
\caption{(Colour online) Panels a and b: A comparison of the behavior of the partial 
excess mixing enthalpies depending on temperature
from simulations of TIP4P/$\varepsilon$--UAM-I model 
(green and magenta lines with symbols for 323.15~K and 278.15~K, respectively) 
and experiments (black lines with similar type of symbols). 
The experimental data are 
from~\cite{iwona} in panel a, and from~\cite{lama} in panel b, respectively. 
In panel b --- $P = 1$~bar, $T = 298.15$~K. Simulations --- red triangles, experimental data --- 
black circles.
Panel c: The pressure dependence of partial excess molar enthalpies from simulations at 298.15~K.
}
\label{fig9}
\end{figure}

In general terms, the computer simulation predictions for $h_{\text{ex}}^{(2)}$ are reasonable
at two values of temperature studied, $T = 278.15$~K and $T = 323.15$~K, as we see from figure~\ref{fig9}a. 
On the other hand, the shape of $h_{\text{ex}}^{(1)}$ from simulations agrees with
experimental trends as well, figure~\ref{fig9}a.
The changes of inclination of $h_{\text{ex}}^{(2)}$ on $X_2$ reflect the peculiarities of the
energetics of mixing of methanol and water. At a low $X_2$, i.e., in the interval of
water-rich mixtures, addition of even a small amount of alcohol species results in
substantial changes of enthalpy (perhaps due to the changes of the hydrogen bonding
network structure). On the other hand, at higher values of $X_2$, water species
are much less bonded between themselves and provide a more comfortable medium for 
the incorporation of methanol molecules. Therefore, enthalpy changes are much less
drastic. A ``crossover'' occurs at $X_2 \approx 0.3$, i.e., it starts at 1 MeOH - 2 H$_2$O
composition. Concerning the absolute values of these partials, we observe a
discrepancy between the simulations and experimental data. 
It is worth mentioning that in  water-methanol mixtures, the magnitude of changes 
of $h_{\text{ex}}^{(i)}$ on $X_2$ is much less drastic, in comparison to ethanol-water 
mixtures, i.e., for the alcohol with a larger hydrophobic tail,  
studied by us very recently~\cite{benavides}.
In summary, the 
TIP4P/$\varepsilon$ - UAM-I model provides a satisfactory
description of the energetic trends of mixing of methanol and water species
under such conditions. 

Difficulties in the exploration of partial properties are illustrated in panel b of Figure 9.
Namely, the partial derivatives of the excess mixing enthalpy are evaluated
by using a restricted set of experimental points. As a result, noisy
curves are obtained. On the other hand, one can
use a more ``chemical engineering procedure'' by the development of 
desirably accurate, analytical polynomial expression for a given property
and following a much easier differentiation. One example of this semi-empirical type of 
approach is the Tait equation of state for water density up to high pressures, 
see e.g.,~\cite{tait}. 
We hope to extend the present observations accumulating more data for enthalpy and to
develop a more successful fitting procedure in a future work.

Our final remarks in this subsection concern the results shown in panel c of figure~\ref{fig9}.
The curves describe how the partial excess enthalpies of methanol and water species
change with pressure at a fixed temperature, $T = 298.15$~K. The shape of each
partial excess enthalpy does not change significantly upon increasing pressure
from 1 bar to 1500 bar. Higher pressure leads to a more negative partials $h_{\text{ex}}^{(1)}$ and
$h_{\text{ex}}^{(2)}$, in close similarity to our observation for the excess mixing
enthalpy in figure~\ref{fig8}c. Apparently, the value for composition, $X_2$, where change of 
slope of the partials occurs is not affected by pressure that seems to be quite
high, $P = 1500$~bar. We are not able to prove whether this tendency remains
at even higher pressures. Moreover, we have not found experimental data that
confirm the predictions of simulations in this aspect. 
In summary, the energetic trends of mixing of alcohol species with water upon 
pressure changes require a more exhaustive laboratory and simulation investigations. 
At the moment, we note that the model predictions are in a qualitative agreement with the
available experimental data.

\subsection{Self-diffusion coefficients depending on temperature and pressure.}

One of the most popular and stringent targets used in the design of the
force fields and in testing  properties for binary mixtures are the
self-diffusion coefficients of species. They can be obtained from the
mean square displacement of particles or from the calculations of the velocity
auto-correlation functions.
We calculate the self-diffusion
coefficients, $D_i$ ($i = 1, 2$), by the former route, via  the Einstein relation,
\begin{equation}
D_i =\frac{1}{6} \lim_{t \rightarrow \infty} \frac{\rd}{\rd t} \vert {\bf r}_i(\tau+t)-{\bf r}_i(\tau)\vert ^2,
\end{equation}
where  $\tau$ denotes the time origin. Default settings of GROMACS were used for the separation of
the time origins. 

\begin{figure}[!t]
\begin{center}
\includegraphics[width=4.48cm,clip]{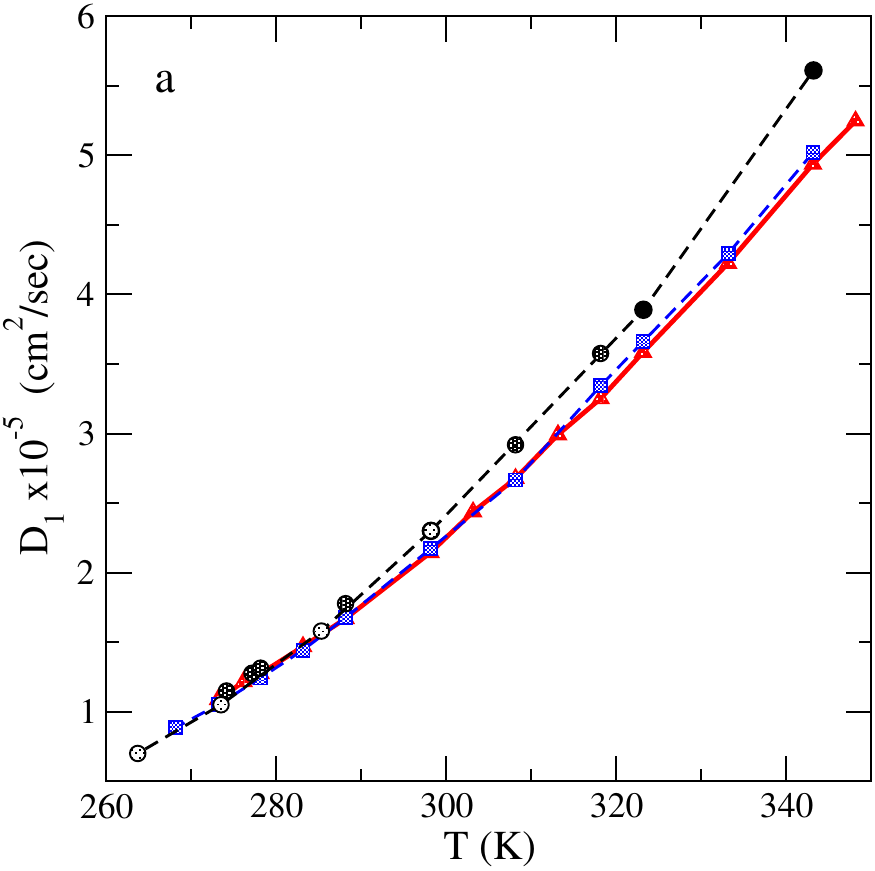}
\includegraphics[width=4.48cm,clip]{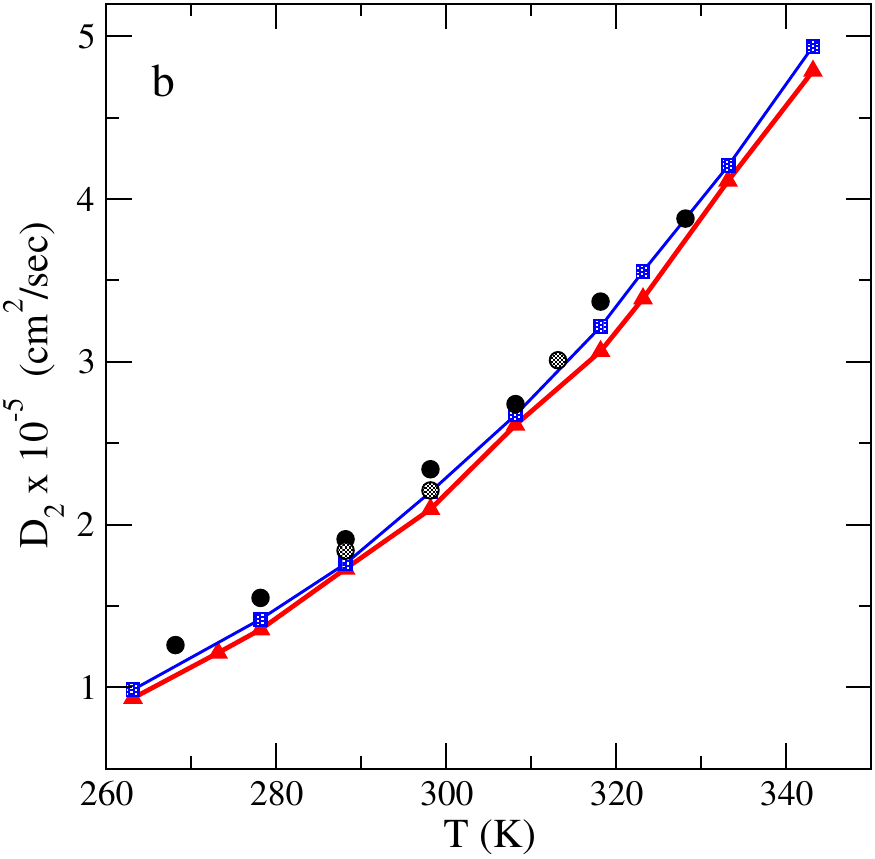}\\
\includegraphics[width=6cm,clip]{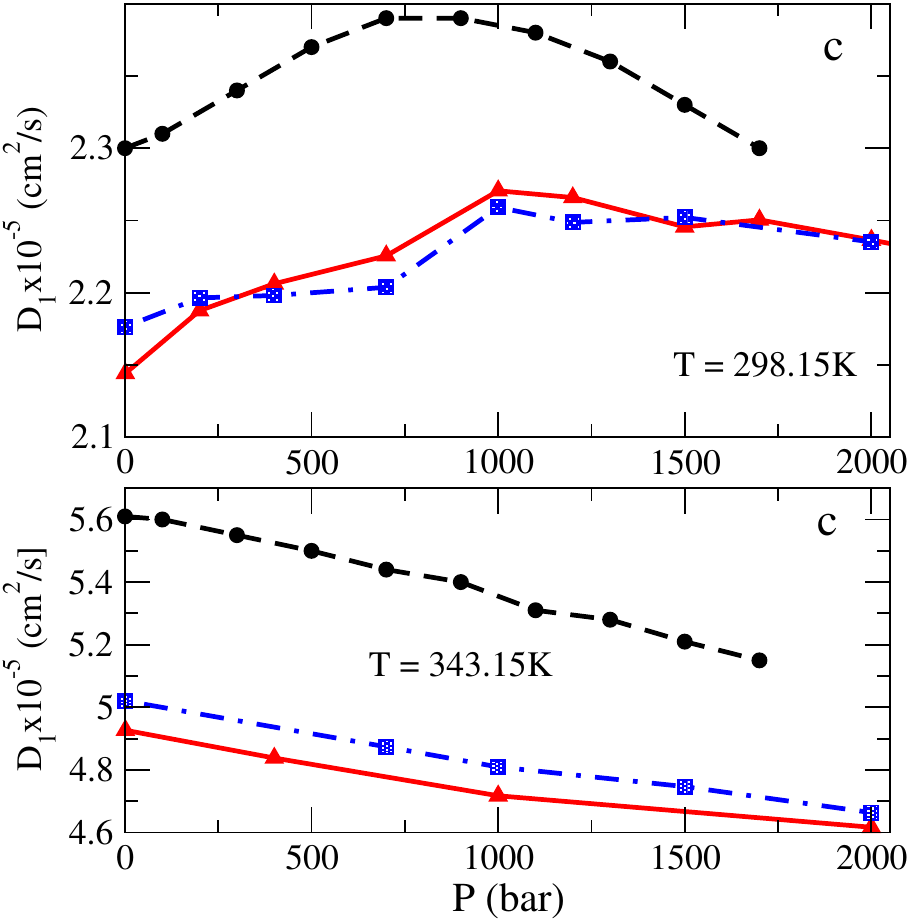}
\includegraphics[width=6.0cm,clip]{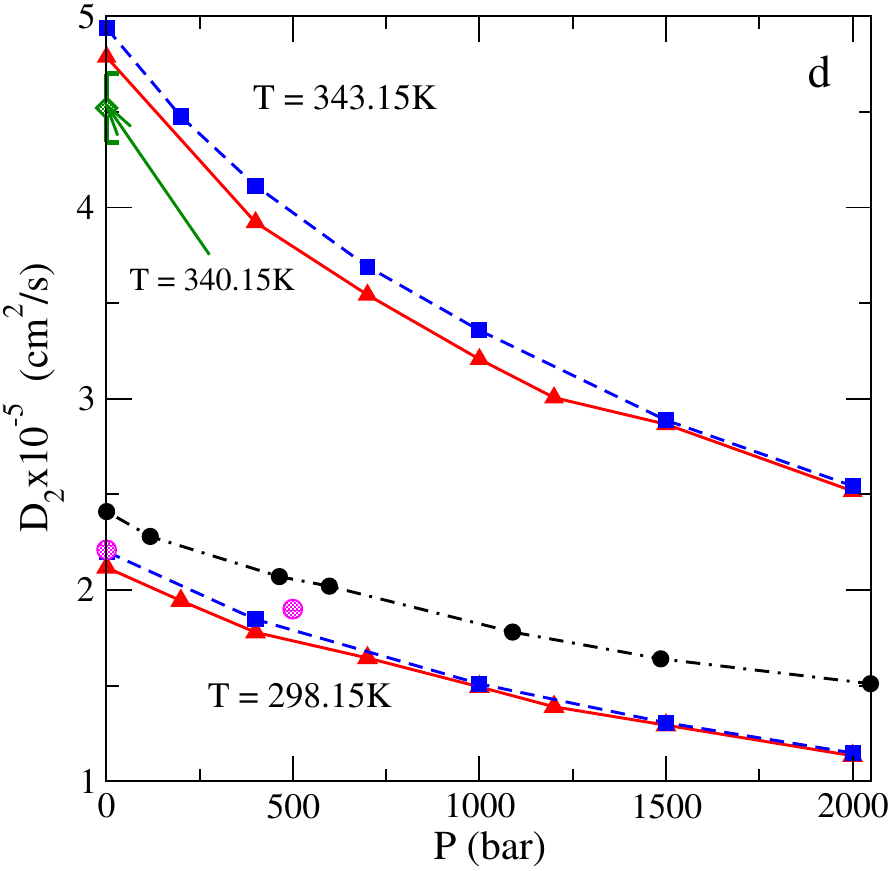}
\end{center}
\caption{(Colour online) Panels a and b: Self-diffusion coefficients of water and methanol depending 
on temperature at ambient pressure. Red triangles and blue squares in panel a
are for TIP4P/$\varepsilon$ and TIP4P/2005 water models, respectively.
The experimental data are from Krynicki et al.~\cite{krinicki},
Mills~\cite{mills}, and Gillen~\cite{gillen},
respectively (circles with decreasing black colour intensity).
In panel b: red triangles and blue squares are for UAM-I and TraPPE methanol
models, respectively. The experimental data are from~\cite{rathbun}
and~\cite{hiraoka}
(circles with decreasing black color intensity).
Panel c: $D_1(P)$ from simulations of TIP4P/$\varepsilon$ and TIP4P/2005 water models
(the nomenclature of lines as in panel a).
The experimental data are from~\cite{krinicki} (black circles)
Panel d: $D_2(P)$  from simulations of UAM-I  and TraPPE methanol models
at different temperatures.
The experimental data are from~\cite{hurle} (black circles with dash-dotted line),
from~\cite{hiraoka} (magenta circles), the green diamond with error bars is
from~\cite{metha}.
The nomenclature of colours and lines for simulation results --- as in panel a.
}
\label{fig10}
\end{figure}

Let us begin from the data for individual species.
The experimental data for the self-diffusion coefficient of water depending on temperature
have been taken from~\cite{krinicki,mills,gillen}, panel a of figure~\ref{fig10}.
Both simulated models, TIP4P/2005 and TIP4P/$\varepsilon$, describe $D_1(T)$ quite
well. Still, one observes that the $D_1(T)$ from simulations is underestimated
in the temperature interval from $T \approx 300$~K up to 343.15~K, in comparison with
experimental data. On the other hand, the UAM-I and TraPPE models for methanol
describe $D_2(T)$ very well in the entire temperature range studied (figure~\ref{fig10}b).
The simulation data in both panels of figure~\ref{fig10} refer to $P = 1$~bar.

Now, let us inspect the dependencies $D_1(P)$ and $D_2(P)$ at a fixed temperature.
For water we observe two types of behavior shown in figure~\ref{fig10}c. At room temperature,
298.15~K, the self-diffusion coefficient, $D_1$,  increases upon increasing pressure
from 1 bar to $P \approx 700$ bar and then decreases with further growth of pressure.
The simulation data, for both water models involved, exhibit similar trends. However,
the maximum of $D_1(P)$ occurs at a higher value of pressure, $P \approx 1000$ bar,
in comparison with the experimental results. The discrepancy between simulations and experiments
is not big in absolute values in the entire interval of pressures. At a higher fixed temperature,
343.15~K, the simulations and experiments show that $D_1(P)$ smoothly decreases with
increasing pressure. This kind of behavior is characteristic of simple liquids.
Thus, in both water models, the effects of hydrogen bonding implicitly hidden in the 
self-diffusion coefficient are weak. On the other hand, at a lower room temperature, 
``a simple fluid''-like behavior is observed solely at high pressures, probably because
the hydrogen bonded network is severely damaged already at $\approx$ 700 bar.

\begin{figure}[!t]
\begin{center}
\includegraphics[width=7.0cm,clip]{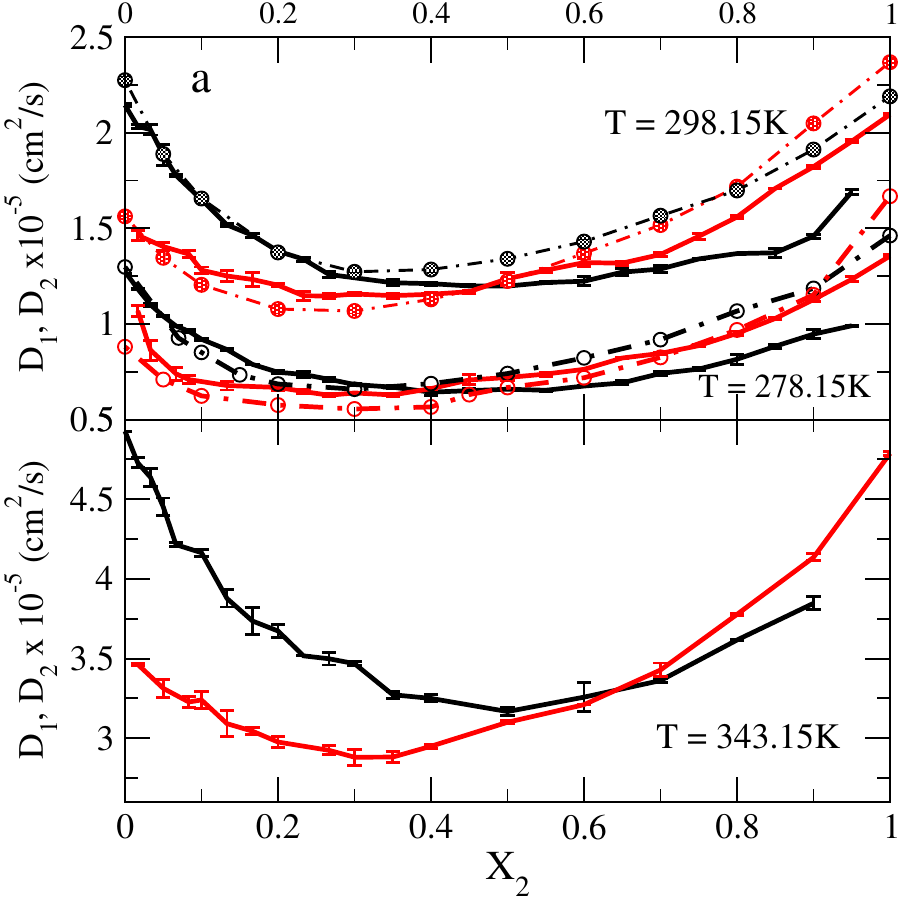}
\includegraphics[width=7.0cm,clip]{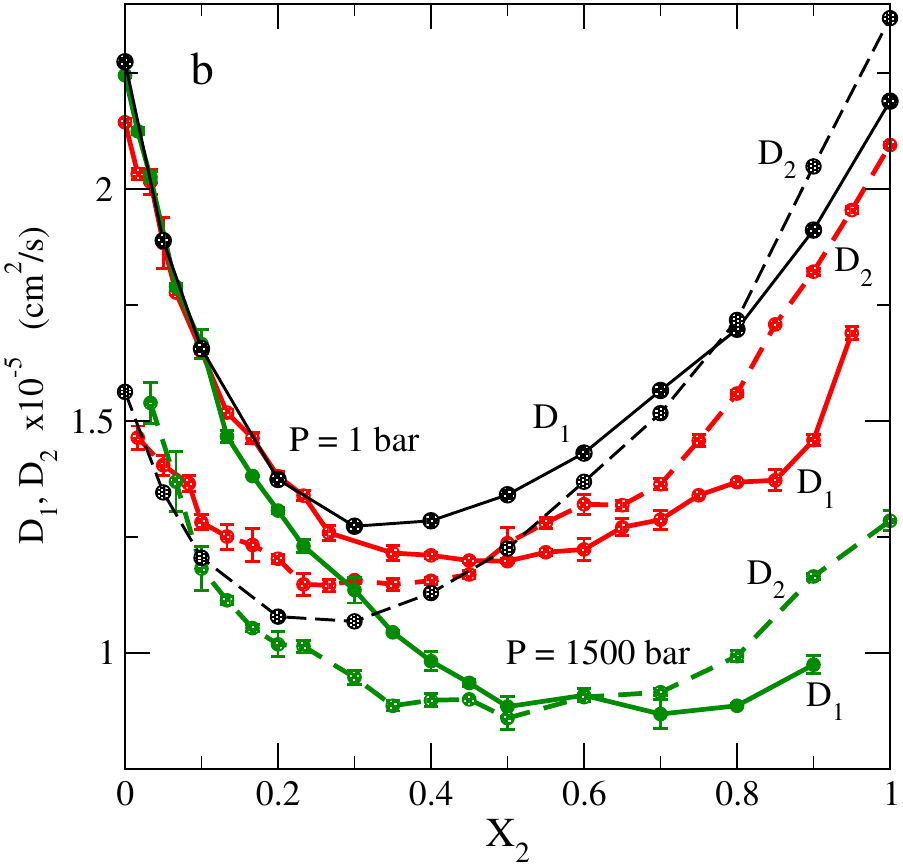}
\end{center}
\caption{(Colour online) Panel a: Composition dependence of the self-diffusion coefficients of water
and methanol at different temperatures at a fixed pressure $P = 1$ bar.
The red and black curves correspond to methanol and water species, respectively.
The experimental data in the upper part of panel a are marked by dash-dotted
lines with circles~\cite{derlacki}.
Panel b: Composition dependence of the self-diffusion coefficients of water
and methanol at different pressures at a fixed temperature 298.15~K.
The solid and dashed lines correspond to water and methanol species, respectively.
Black lines with circles --- experimental data~\cite{derlacki} at $P = 1$~bar.
Red lines --- 1 bar; green lines --- 1500 bar.
}
\label{fig11}
\end{figure}

In contrast to water, the self-diffusion coefficient of methanol, $D_2(P)$ monotonously 
decreases with increasing pressure at both fixed temperatures, 298.15~K and 343.15~K 
(figure~\ref{fig10}d). Thus, the effects of bonding between methanol molecules are not strong
under conditions of our study. Concerning the agreement between the predictions of two
methanol models in question and  experimental data, it can be termed as reasonable at 
298.15~K. However, at a higher temperature of our interest, we have not found systematic
data. Moreover, the point at $P = 1$ bar from~\cite{metha} at 340.15~K is not definite.
Other experimental techniques yield even a higher value for $D_2(P)$ at this pressure.

Now, we would like to turn our attention to water-mixture and explore the
evolution of $D_i(X_2)$, $i = 1,2$,  with temperature and pressure. The simulation results and
available experimental data are given in figure~\ref{fig11}.
From panel a of this figure, we learn that the composition dependence of the
self-diffusion coefficients of both species from simulations is in qualitative 
agreement with the experimental data from~\cite{derlacki} at 278.15~K and
298.15~K. The diffusion in water-rich mixtures is better described than in mixtures
with higher concentration of alcohol species. Concerning the changes of magnitude
of $D_1$ and $D_2$ with increasing temperature, we observe that the trends
captured by simulated models are correct, as it follows from the inspection of
available experimental data. However, the crossover composition value 
between the regions where $D_1 > D_2$ and where $D_1 < D_2$ is not accurately 
reproduced by the model under study. Similar observations are valid for the
behavior of $D_1$ and $D_2$ upon increasing pressure. The trends of changes
concerning the magnitudes of the self-diffusion coefficients are
intuitively correct. However, we were unable to find laboratory data confirming the
simulation predictions.

\subsection{Summary and conclusions}

To conclude, we report our fresh results concerning the trends of behavior
of the mixing properties and self-diffusion coefficients of water-methanol liquid
mixtures. They are studied dependent on the temperature, pressure
and composition by using isobaric-isothermal molecular dynamics simulations.
This study is just the first part of the ampler project that implies an investigation
of the microscopic structure in terms of various pair distribution functions, 
coordination numbers and hydrogen bonding network evolution.
Besides, we contemplate to explore the dynamic behavior of the mixtures in question
in terms of viscosity and various relaxation times, dielectric 
and interfacial properties. Progress of research along these lines will be
reported in separate publications.
The mixtures in question are considered by using the recently proposed parametrization
within the TIP4P$\varepsilon$--UAM-I model. In the case of pure components,
we have involved the TIP4P/2005 water model as well as the TraPPE methanol model.
However, in the case of mixtures, solely a single combination of water and
methanol modelling has been considered for the moment. Evaluation of the performance
of different combinations of models is postponed to future work.
The motivation of using the present version of the model for mixtures has two reasons.
On the one hand, a possible extension to other monohydric alcohols aqueous solutions depending
on temperature and pressure is ensured, as we showed in part in our 
recent work~\cite{bermudez}. On the other hand, in order to put closer
the simulation outputs and several laboratory works by using dielectric spectroscopy,
one would need a confiable description of the dielectric constant, at least.
Temperature dependence of the dielectric constant of water 
is satisfactory within the TIP4P$\varepsilon$ modelling as shown in~\cite{fuentes}.
Thus, the present modelling is quite promising for water-alcohol mixtures. 

Most interesting and presumably important findings of the present work are
as following. Concerning the temperature trends of the apparent molar volume
of methanol species, we have shown that the minimum of $V_{\phi}^{(2)}$
disappears with increasing temperature in the interval between 323.15~K and 343.15~K.
Moreover, the dependence of the most favourable contraction of the 
volume on temperature is reproduced by simulations as well.
Pressure trends of this property exhibit an interesting evolution as well.
Namely, the minimum of $V_{\phi}^{(2)}$ is observed at all pressures, 
upon increasing it up to 4 kbar. However, at a high pressure close to 4 kbar, we observe 
a crossover for the mixing volume, $\Delta V_{\text{mix}}$, from negative to positive
values. This implies an expansion of the volume in water-rich mixtures, in contrast to the 
contraction observed in mixtures with a higher amount of alcohol (methanol).
Consequently, the reduced apparent molar volume of methanol species becomes
a bit larger than 1 at low $X_2$ values.
This trend of behavior is definitely related to the evolution of the hydrogen bonds
network upon adding alcohol to water. A set of issues related to this finding,
or say its interpretation, will be discussed elsewhere in the studies of microscopic
structure. One more interesting trend, resulting from our calculations,  is the
change of the excess mixing enthalpy upon increasing temperature. Namely,
the computer simulation results of the model evidence the changes from the exothermic to endothermic
mixing in the temperature interval from 423.15~K to 473.15~K. In qualitative terms,
this finding is in agreement with the experimental data.

Finally, we have performed a rather detailed investigation of the behavior of self-diffusion coefficients depending on temperature and pressure for individual species and
for the mixtures in the entire range of composition. We observed that the dependence of
the self-diffusion coefficient of water on pressure is different at room temperature,
298.15~K, and at a higher temperature, 343.15~K. Namely, the $D_1(P)$ (1~refers to water)
increases if pressure grows from 1 bar to $\approx$ 1 kbar and then
decreases with the further increasing pressure, at $P > 1$ kbar, if $T = 298.15$~K.
At a high temperature, $T = 343.15$~K, $D_1(P)$ monotonously decreases upon increasing pressure.
These changes of 
the behavior for $D_1(P)$ are related to the evolution of hydrogen bonding structure
of water. On the other hand, the self-diffusion coefficient of methanol decreases
with increasing pressure in the entire interval of temperature and pressure studied,
as expected. These trends are then manifested in the evolution of $D_1(X_2)$ and $D_2(X_2)$
at different thermodynamic states, determined by $T$ and $P$. The accuracy of these predictions
is difficult to evaluate at the moment because the experimental data are scarce.

\section*{Acknowledegments}
The authors acknowledge support by CONAHCyT of Mexico under the grant  CBF2023-2024-2725.
Helpful discussions with Dr. Laszlo Pusztai stimulated this work. Technical
support of Magdalena Aguilar at Institute of Chemistry of the UNAM is acknowleged as well.
We express our gratitude to all of them.

\ukrainianpart

\title
{Про вплив температури, тиску і складу водно-метанольних сумішей на їх властивості. I. Густина, надлишковий об’єм змішування та ентальпія, а також коефіцієнти самодифузії, отримані методом молекулярної динаміки.
}

\author{М. Круз Санчес\refaddr{label1},
	В. Трехос Монтоя\refaddr{label1}, 
	О. Пізіо\refaddr{label2}}
\addresses{
	\addr{label1}
	Хімічний факультет Автономного університету Метрополітана-Істапалапа, просп. Сан Рафаель Атлікско 186, 09340, CDMX, Мехіко
	\addr{label2}  Інститут Хімії, Національний Автономний Університет Мексики, Мехіко
}

\makeukrtitle

\begin{abstract}
	\tolerance=3000%
З використанням комп’ютерного моделювання в рамках ізобарно-ізотермічної молекулярної динаміки, досліджено
залежність деяких основних властивостей модельних сумішей води та метанолу від температури, тиску та хімічного складу.  Основна увага зосереджена на неполяризаційній моделі UAM-I-EW 
об'єднаного	атома метанолу, яка нещодавно була параметризована у V. Garcia-Melgarejo та ін. [~J. Mol. Liq., 2021, \textbf{323}, 114576], у поєднанні з моделлю води TIP4P/$\varepsilon$. У перспективі, модель метанолу дозволяє зручне узагальнення теорії на випадок інших одноатомних спиртів, змішаних з водою. Описано поведінку густини, надлишкового об'єму змішування та ентальпії; інтерпретуються властивості часткового змішування. Крім того, досліджено тенденції поведінки коефіцієнтів самодифузії частинок суміші. Якість модельних передбачень критично оцінюється шляхом детального порівняння з експериментальними результатами. Отримані результати є новими та дають змогу зрозуміти поведінку подібних сумішей при різних температурах та високих тисках. Обговорюється вдосконалення моделювання, необхідне для подальших досліджень.
	\keywords моделювання методом молекулярної динаміки,  суміші води та метанолу, парціальний молярний об'єм, надлишкова ентальпія, коефіцієнти самодифузії
	
\end{abstract}
\end{document}